\documentclass[12pt]{article}
\usepackage{amsfonts}
\usepackage{geometry}
\usepackage{verbatim}
\usepackage{graphics}
\usepackage{epsfig}
\usepackage{amsmath}
\usepackage{amssymb}
\usepackage{amsthm}
\usepackage{lscape}
\usepackage{setspace}
\usepackage[bottom]{footmisc}
\usepackage{endnotes}
\usepackage{setspace}
\usepackage{xcolor,url}
\usepackage[round]{natbib}
\usepackage{booktabs}
\usepackage{subcaption}

\def\tfrac#1/#2{\leavevmode
        \kern.1em \raise .5ex \hbox{\the\scriptfont0 #1}%
        \kern-.1em $/$%
        \kern-.15em \lower .25ex \hbox{\the\scriptfont0 #2}}
\def\makeactive#1{\catcode`#1=\active \ignorespaces}
\def\alwaysspace{\hglue\fontdimen2\the\font \relax}%
{\makeactive\^^M \makeactive\ %
\gdef\obeywhitespace{%
\makeactive\^^M\def^^M{\par\indent}%
\makeactive\ \let =\alwaysspace}}%
\def\bkR{{\rm I\kern-.17em R}}
\def\uniset{{\rm 1\kern-.40em 1}}
\newenvironment{stdarray}{\[  \left\{ \begin{array}{lcl}}{\end{array} \right. \]}
\newcommand{\bstd}{\begin{stdarray}}
\newcommand{\estd}{\end{stdarray}}
\newcommand{\beq}{\begin{equation}}
\newcommand{\eeq}{\end{equation}}
\newcommand{\bc}{\begin{center}}
\newcommand{\ec}{\end{center}}
\graphicspath{{../../Plots/}{../../Plots/NN/}{../../Plots/RF/}{../../Plots/GBT/}}
\DeclareMathOperator{\cov}{cov}

\input{tcilatex}
\begin{document}

\title{{\Large \textbf{Testing for Asymmetric Information in Insurance\\
with Deep Learning} \thanks{%
We thank Simon Lee and Vira Semenova for their very helpful comments and
suggestions. We are grateful to Marc Maliar for his help with writing the
TensorFlow code for estimation and testing, and to Xiangru Li for excellent
research assistance.}}}
\author{\textbf{Serguei Maliar}\thanks{%
Santa Clara University.} \and \textbf{Bernard Salani\'e}\thanks{%
Department of Economics, Columbia University, bsalanie@columbia.edu.} }
\date{\today}
\maketitle

\begin{abstract}

The positive correlation test for asymmetric information developed by \cite%
{chiappori_testing_2000} has been applied in many insurance markets. Most of
the literature focuses on the special case of constant correlation; it also
relies on restrictive parametric specifications for the choice of coverage
and the occurrence of claims. We relax these restrictions by estimating
conditional covariances and correlations using deep learning methods.{\ We
test the positive }correlation property by using {the intersection test of 
\cite{chernozhukov_intersection_2013} and the {\textquotedblleft sorted
groups\textquotedblright \ test of \cite{chernodemirerduflofv23}. }Our
results confirm earlier findings that the correlation between risk and
coverage is small. Random forests and gradient boosting trees produce
similar results to neural networks.}

\end{abstract}

\onehalfspacing

\thispagestyle{empty}

\clearpage

\setcounter{page}{1}

\section{Introduction}
In insurance markets, the 
\emph{positive correlation property} (PCP) states that   insurees'
choices of coverage should be positively related with ex-post measures of
their risk, such as the occurrence and severity of claims. This can be
either because of adverse selection (i.e., riskier insurees self-select into
contracts with higher coverage, as in \citet{rothschildstiglitz76}), or
because of moral hazard (i.e., a higher coverage discourages prevention
effort). \cite{cjss:2006} proved that the PCP obtains quite generally in
models of competitive equilibrium in insurance markets. 

\cite{chiappori_testing_2000} used the PCP to propose a test for asymmetric
information in insurance; and they applied it  to French
car insurance data. In its simplest form, the PCP states that the
conditional correlation of coverage and ex-post risk should be positive, for
all values of the vector of covariates that are observed both by the insurer
and the insuree\footnote{%
As explained in \cite{cjss:2006} and \cite{dionne_asymmetric_2013}, the
precise statement of the positive correlation property needs to be adapted
in more general models.}. Suppose that coverage is treated as a binary
choice (i.e.\ minimal versus comprehensive); and that ex-post risk is also
binary (e.g.\ whether or not the insuree filed a claim). %
\citet{chiappori_testing_2000} argued that under the PCP, the residuals of
two binary choice models for coverage and ex-post risk should be positively
correlated, if all public covariates are controlled for. Hence, they tested
the hypothesis that the correlation of the generalized residuals of two
univariate probits was zero; they also estimated and tested for the
correlation of the generalized residuals of a bivariate probit.

To their surprise, \citet{chiappori_testing_2000} found no statistical
evidence for the PCP with any of these tests in the French car insurance
data they were using: the correlation of coverage and risk was close to zero.%
\footnote{%
This is not a universal finding; while the PCP was not documented with car
insurance data, there is evidence for the PCP in some other markets. Our
purpose is not to discuss the PCP evidence, \textit{but to explore how the
outcome of the PCP test depends on design and implementation of the testing
procedure.}}  This remarkable fact implies that essentially
all relevant information is contained in publicly observed covariates.

However, the analysis of \citet{chiappori_testing_2000} relied on
restrictive specifications that may have limited power to detect the PCP. 
They only 
included about fifty regressors in their probit regressions, which was 
a very small subset of the covariates they could have constructed by
interacting the covariates. Moreover, their bivariate probit
model was built on a simplifying assumption of constant correlation. 
\footnote{The test procedures in \citet{chiappori_testing_2000} have been extended in
several directions. \citet{KimEtAlJRI09} showed how the probit for coverage
can be replaced by an ordered multinomial choice model when more than two
types of contracts are available. \citet{chiappori_testing_2000}  had also proposed a
fairly basic 
nonparametric test. Following ideas in 
\citet{DionneGV01,
DionneGV06}, \citet{SuSpindlerJBES13} and \citet{SpindlerGRIR14} used a
more powerful nonparametric test of conditional independence. However, this is only
practical when there are no more than three continuous covariates and a
small number of discrete covariates.}.

The extraordinary development  in machine learning methods in the past decade
 suggests revisiting some  seemingly well-established empirical findings.  
Our goal here is to show how these  methods can be
implemented in the context of the PCP, and to check whether  more
powerful testing methods can alter the conclusions of %
\citet{chiappori_testing_2000} about its quantitative unimportance.

Ideally, one would want to do two things: estimate flexibly the conditional
correlation of risk and coverage for any given values of the covariates, and
test that this conditional correlation is positive for all values in a given
subset (e.g. for all male, 40- to 45-year old drivers who use a 5-year old
car). We show that both goal can be achieved by combining the flexibility of
machine learning methods and standard econometric tests. From a growing
catalog of machine learning methods on the market, we choose deep learning
as our main estimation method because of its popularity and its remarkable
success in many applications.

For the sake of comparison, we use the same car insurance data as in %
\citet{chiappori_testing_2000}. The dataset contains only 6,333
observations, a typical size for many microeconometric applications. This
allows us to explore the effectiveness of deep learning techniques in such
settings. Each observation includes information on the car (brand, model,
age, power, \ldots ) and the client's demographics (age, profession,
residence,\ldots ); we use these variables and their interactions to
construct the covariates $x$. We summarize coverage $c$ and ex-post risk $r $
by two binary variables. Let $X$ denote all publicly observed covariates.
We denote $p_{jk}(x)$ the probability that $c=j$ and $r=k$ given $X=x$, for $%
j,k=0,1$. Finally, we let $p(x)=p_{10}(x)+p_{11}(x)$ (resp.\ $%
q(x)=p_{01}(x)+p_{11}(x)$) denote the probability that $c=1$ (resp.\ that $%
r=1$) conditional on $X=x$. The probability $p(x)$ is the conditional choice
probability of the higher coverage, and $q(x)$ is the conditional
probability of an at-fault claim.

The most basic form of the positive correlation property states that for all
values of $x$, the covariance of $c$ and $r$ conditional on $X=x$ is
non-negative:%
\begin{equation*}
C(x)\equiv \cov(c,r|X=x)=p_{11}(x)-p(x)q(x)\geq 0.
\end{equation*}%
A shortcoming of the covariance is that its value is not easily
interpretable. We therefore also state the positive correlation property in
terms of the correlation coefficient: 
\begin{equation*}
\rho(x)\equiv \frac{C(x)}{\sqrt{p(x)(1-p(x))q(x)(1-q(x))}}\geq 0.
\end{equation*}

Estimating the covariance and correlation functions for given values of the
covariates $X=x$ requires estimating the $p_{jk}$ probabilities flexibly.
This is not a simple task, as many interactions between the covariates can
have explanatory power. It is notoriously hard, for instance, to model the
risk $q(x)$ parsimoniously. It is even more difficult to test that $\rho $
is non-negative over a subset of covariates: interesting subsets typically
are very large, leading to a multiple testing problem where the
distributions of the estimated covariances or correlations for different $x$
are not independent.

We implement three approaches to apply machine learning to testing for the PCP. Our first
test relies on a feedforward neural network to predict the conditional
probabilities $p_{jk}(x)$: in the terminology of this field, this is a 4-way
classification problem. A potential complication we face when testing the
positive correlation property is that the neural network estimates $\hat{p}%
_{jk}(x)$ have a relatively slow rate of convergence and act as nuisance
parameters. As it turns out, the covariance function has a nice double
robustness property; the presence of these nuisance parameters is not an
issue. On the other hand, the presence of such nuisance parameters does
complicate inferences about the correlation function.

To remedy this, we use the double-debiasing method of~\cite%
{chernozhukov_doubledebiased_2018}, which extends the idea of Neyman
orthogonalization to a broad range of models and estimation methods. We
combine this double-debiasing method with results from \cite%
{SemenovaCherno:EJ2021} to obtain consistent and asymptotically normal
estimators of the average values of the covariance and correlation function
within groups of observations. We then use the intersection tests developed
in (\cite{chernozhukov_intersection_2013}) to test the positive correlation
property for a variety of groups of observations. We test, for instance,
that the correlation is positive on average for all modalities of the
\textquotedblleft age of the car\textquotedblright \ variable.

We find that the neural network predicts the purchase of coverage
considerably better than it does the accident occurrence. This is not that
surprising: at-fault claims are relatively low-probability events and
insurers know that they are hard to predict. The range of the estimated
covariance function lies mostly in the interval $[-0.01,0.01]$. While the
correlation function has a larger range, it narrows considerably when we
double-debias it and we average within groups. Our intersection tests show
that for any of our eight covariates, we can reject the hypothesis that the
correlations are positive on average for all values of its modalities. On
the other hand, we cannot reject the hypothesis that these average
correlations are larger (i.e., less negative) than a small negative number
like $-0.05$. In the end, we obtain a 95\% confidence interval for the range
of these average correlations that is confined to a narrow interval around
zero.

Our second method relies on the \textquotedblleft sorted
groups\textquotedblright \ approach of~\cite{chernodemirerduflofv23}.
 We first cross-fit the neural network model that we selected in our first approach. We 
 then allocate observations  into groups sorted
by the value of the predicted covariance or correlation.  To maximize the power of the test of the
PCP, we focus on the observations for which the estimated correlation is
smallest in algebraic terms.  We find here again 
that only very small values of the covariance and correlation are consistent
with the data, and we find no significant evidence for  the PCP.

Finally, we run two variants of our first approach in which we replace the neural network with 
two other popular machine learning methods---random
forests and gradient-boosted trees. 
While these two methods put different weights on the various covariates, the
results of the intersection tests are similar to those obtained by using our
baseline deep learning method.

To summarize:  machine learning methods  do not alter the qualitative conclusion of %
\citet{chiappori_testing_2000} that the correlation of coverage and risk is
essentially zero. Even more remarkably, we could not find evidence for a positive
correlation  in any reasonably-sized 
subpopulation. While  adverse selection and moral hazard are clearly important phenomena
in many markets, they do not seem to play much of a role in this one.

The rest of the paper is as follows. Section~\ref{sec:data} describes the
data used for estimation. Our analysis has three steps: In Section~\ref%
{sec:nn}, we fit a neural network to classify insurees into the four
alternatives $c,r\in \left \{ 0,1\right \} \times \left \{ 0,1\right \} $. In
Section~\ref{sec:covcorr}, we use the double-debiasing methods of \cite%
{chernozhukov_doubledebiased_2018} to correct the correlation function. In
Section~\ref{sec:PCP} we test the positive correlation property by running
the intersection test of \cite{chernozhukov_intersection_2013}. Finally,
Section~\ref{sec:RFandGBT} compares our results with those obtained using
methods based on decision trees.

\section{The Data}\label{sec:data}

\citet{chiappori_testing_2000} obtained their data from the French
federation of insurance companies FFSA, which ran a survey of automobile
insurance in 1989. They selected a subsample that only includes
\textquotedblleft young\textquotedblright \ drivers---insurees who obtained
their driver's licence within the past three years. We focus on an even
narrower subsample of insurees whose driving license is one year old at
most. Since these individuals have no previous driving history, there is no
concern about the impact of experience rating on driving behavior; it also
reduces the unobserved heterogeneity in the sample.

Our selection leaves us with a sample of 6,333 observations. The data on
each insuree and car are quite rich. Each observation includes information
on the car (brand, model, age, power, \ldots ), the client's demographics
(age, profession, residence,\ldots ), the type of contract, and the claim
record. As in~\citet{chiappori_testing_2000}, we code the type of contract
and the claim record as binary.

The data also record if the insurance contracts covered all or part of the
year. Only about 40\% of insurees were insured throughout the year, and
roughly 15\% were covered for less than two months. Like in %
\citet{chiappori_testing_2000}, we use sampling weights $w$ that represent
the number of days that the insuree was insured during the year. We indicate
sampling weights with $w$ subscripts when needed.

In automobile insurance, the main distinction between contracts is whether
they only include third-party (liability) coverage---which is compulsory in
France---or whether they also cover damages that the insuree caused to
his/her own car. We call the latter \textquotedblleft comprehensive
coverage\textquotedblright , and we will neglect variations within this
class, such as the amount of the deductible. Since the difference between
third-party and comprehensive coverage only matters when the insuree is at
fault, we define our claim variable accordingly. This results in the two
binary variables $c,r\in \left \{ 0,1\right \}$:

\begin{itemize}
\item $c=1$ if the insuree opted for comprehensive coverage

\item $r=1$ if the insuree filed at least one at-fault claim.
\end{itemize}

We also define four indicator variables as $y_{jk}=1$ if $(c=j\mbox{ and }%
r=k)$ for $j,k=0,1$. Table~\ref{tab:crobs} classifies the 6,333 observations
by the four indicators.

\begin{table}[th]
\centering%
\begin{tabular}{l|cccr}
 Event & $c$ & $r$ & Alternatives & Number of observations \\ 
\midrule Third-party, no accident & 0 & 0 & $y_{00}=1$ & $3,696$ \\ 
Third-party, accident & 0 & 1 & $y_{01}=1$ & $302$ \\ 
Comprehensive, no accident & 1 & 0 & $y_{10}=1$ & $2,203$ \\ 
Comprehensive, accident & 1 & 1 & $y_{11}=1$ & $132$ \\ 
\midrule 
\multicolumn{4}{l}{Total} & $6,333$ \\ 
\end{tabular}%
%
\caption{Observed Classification}
\label{tab:crobs}
\end{table}

We use the same set of covariates $X$ as~\citet{chiappori_testing_2000};
they are created from the eight variables that insurers identified as being
the most important. We have up to 28,800 categories of insurees: nine age
categories, eight professions, four types of use, ten regions, five
rural-to-urban codes, and gender; and 72 car categories: six categories for
the performance of the car, and twelve for its age. Combining them would
yield more than 2 million dummy variables, a number that dwarfs the sample
size. Even if we had a much larger dataset, there are many more variables in
the data. This is a clear-cut \textquotedblleft $p\gg n$\textquotedblright \
case, which calls for model selection.

\section{The Deep Learning Model}

\label{sec:nn} We fit the 4-way probabilities $p_{jk}(x)=\Pr
(c=j,r=k|X=x)$ for $j,k=0,1$ with a neural network. The values of the eight
covariates $x$ enter the input layer. The neural network has $D$ hidden
layers, each with $W$ neurons; the last hidden layer feeds into a 4-node
output layer which uses a \textquotedblleft softmax\textquotedblright \
function to generate the probabilities $p_{jk}(x)$. We train the neural
network using the Adam optimizer \citep{KingmaAdam:2017} to minimize the
cross-entropy loss function adjusted by the sampling weights. For a sample
of observations $(x_{i},c_{i},r_{i},w_{i})_{i\in I}$, the loss is 
\begin{equation*}
L=\sum_{i\in I}w_{i}\sum_{j,k=0,1}\uniset(c_{i}=j,r_{i}=k)\log p_{jk}(x_{i}).
\end{equation*}%
Our code relies on the Python package Keras \citep{CholletKeras:2021} with
the TensorFlow backend \citep{AbadiTensorflow:2015}.

The class of neural networks we consider has $P\equiv n_{X}W+(D-1)W^{2}+3W$
parameters for $D>0$, where $n_{X}$ is the number of covariates in the input
layer\footnote{%
For $D=0$, the number of parameters is $3n_{X}$.}. Adding up the number of
categories (minus one category per variable) of our eight covariates, plus
the constant, gives us $n_{X}=49$ and results in $P=W(2+(D-1)W)$ parameters.
To illustrate, a very modest neural network with $D=2$ hidden layers of $%
W=16 $ neurons each has $P=1,088$ parameters.

With such a large number of parameters, overfitting is an obvious concern.
To guard against it, we resort to cross-validation: we use a validation
sample to decide when to stop learning. In addition, we randomly drop out
some of the neurons in the hidden layers during training. The idea of
dropout regularization, pioneered by \citet{SrivastavaDropout:2014}, is that
the network will learn to compensate for the missing neurons, and that the
resulting model will be more robust to overfitting. The fraction $d$ of
neurons that are dropped out is called the dropout rate. Dropout allows us
to use a more powerful network with a larger number of neurons. The optimal
dropout rate typically increases with the number of neurons in each hidden
layer, as a higher dropout rate is needed to deal with higher overfitting.

\subsection{Hyperoptimizing the Neural Network}

\label{sub:hyperopt}

The neural network we consider has many hyperparameters: its depth $D$ and
width $W$, the size of the validation sample, the size of the mini-batches,
the choice of optimizer, etc.\ We decided to optimize over the depth and
width, and also over the dropout rate $d$.


Since our sample size is relatively small, we only fit smallish neural
networks that vary in:

\begin{itemize}
\item the number $D$ of hidden layers: we tried 0, 1, 2 and 3 hidden layers;

\item the number $W$ of neurons in each layer: tried 8, 16, and 24 neurons
in each layer;

\item the dropout rate $d$: we tried values from $0$ (no dropout) to $0.8$.
\end{itemize}

To optimize in hyperparameter space, we split the data into training,
validation and testing sets, representing 70\%, 15\%, and 15\% of the data,
respectively. For each 3-uple of values of $D$, $W$ and $d$, we fit the
neural network on the training sample and we use the validation sample to
stop training soon after the loss on the validation sample stops decreasing.
To discard local minima, we keep training the network for a small number of
epochs to see if the validation loss keeps increasing\footnote{%
The parameter that controls the number of periods that we wait is called
\textquotedblleft patience\textquotedblright ; we set it at 10 epochs.}. We
then store the parameters of the neural network that correspond to the epoch
with the smallest validation loss. Then we measure the loss on the test
sample.

After fitting all of these models, we select the combination of the three
hyperparameters that leads to the smallest loss on the test sample\footnote{%
Fitting a neural network involves a choice of initial values for the
weights. TensorFlow uses a well-tested procedure that injects some
randomness into the process. As a result, different runs can select slightly
different neural networks. We find that this had almost no impact on the
results of our final tests of the positive correlation property.}. The
hyperoptimized model has $D=2$ hidden layers; $W=16$ neurons in each layer;
and a dropout rate $d=0.1$. The hyperoptimizing procedure took 150 seconds
on a Mac Studio. The resulting neural network has 1,524 parameters.

We also fit a 2-way classification model for $c$ and another for $r$; we
hyperoptimized them in a similar way. The neural network for $c$ (the choice
of coverage) has no hidden layer and dropout = $0.7$. The model for $r$ (the occurrence
of a claim) has 2 hidden layers; 8 neurons in each layer; and dropout = $0.4$.

\begin{table}[htbp]
    \centering\begin{tabular}{l|rrr}
        Variables & Constant & Probit & Neural network \\ \toprule
        $(c,r)$ & $0.963$ & $0.629$ & $0.441$ \\
        $c$  & $0.673$ & $0.385$ & $0.256$ \\
        $r$ & $0.283$ & $0.244$ & $0.180$ 
    \end{tabular}
    \caption{Comparing Losses}
    \label{tab:losses}
\end{table}

These loss values can be compared with those of probit models
that use the same set of regressors as in \cite{chiappori_testing_2000}, as well as to ``constant'' models
that use no regressors. As Table~\ref{tab:losses}  shows, our deep learners massively outperform the probits of
\cite{chiappori_testing_2000}. 
Since the loss is just minus the average log-likelihood per observation,  these 
 improvements are quite large. For instance, the $0.188$ gain in average log-likelihood  from   bivariate probit 
 to the bivariate deep learner yields a likelihood ratio statistic of $2\times 6,333 \times 0.188=2381.2$, for
 an additional $1,524 - (2\times 48 +1)=1,425$ parameters.  The corresponding $p$-value is minuscule\footnote{This
 is only meant to be illustrative; it is not clear that one can use asymptotic approximations with such
 a large number of parameters.}.

We also applied our procedure to the sampling weights. We use an activation
function that takes into account the truncation of weights in $[0,1]$ and a
loss function that allows for a mass point at $w=1$. Not surprisingly, the
selected neural network is rather sparse: it has no hidden layers ($D=0$)
and a dropout rate $d=0.5$. The whole hyperoptimizing procedure took 130
seconds. The hyperoptimized model has a loss of $0.098$ on the test sample,
while a constant model has a test loss of $0.114$,  which is only 15\% larger
than the model.

\subsection{Estimated probabilities and weights}

\label{sub:estimates}



\begin{figure}[tbp]
\begin{center}
\includegraphics{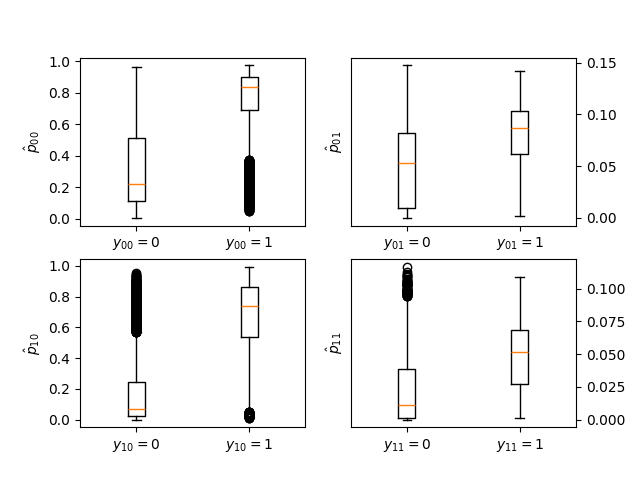}
\end{center}
\caption{Fitting the $y_{jk}$ variables}
\label{fig:y_fits_NN}
\end{figure}
\begin{figure}[tbp]
\begin{center}
\includegraphics[scale=0.7]{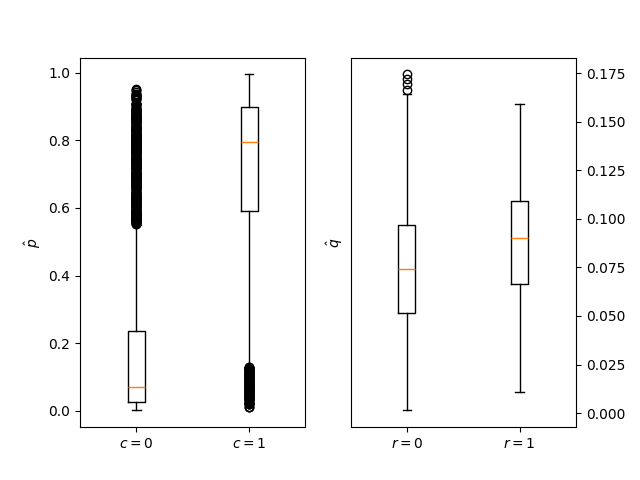}
\end{center}
\caption{Fitting the $c$ and $r$ variables}
\label{fig:pqhat_fits_NN}
\end{figure}

Figure~\ref{fig:y_fits_NN} plots the estimated probabilities $\hat{p}_{jk}$,
both when the corresponding $y_{jk}$ is~0 and when it is~1. The dashed red
line represents the mean of $y_{jk}$ in the sample. A perfect fit would have 
$\hat{p}_{jk}=y_{jk}$ within each panel. It is clear that the left column
performs better than the right column in this respect. It is not that
surprising as there are more than 15 times as many observations with $r=0$
as with $r=1$: the neural network puts a large weight on fitting these
observations. Figure~\ref{fig:pqhat_fits_NN} shows that the model can
predict the purchase of coverage considerably better than the accident
occurrence. This is a common finding with insurance data. It is more
surprising that the neural network overestimates the probability of a claim
to the extent shown in the right panel. This seems to be a side effect of
its efforts to fit the choice of contract (the variable $c$). For
comparison, Figure~\ref{fig:pqhat_fits_NN} shows the results obtained with
the 2-way classification neural networks for $c$ and for $r$.

\begin{figure}[tbp]
\begin{center}
\includegraphics[scale=0.7]{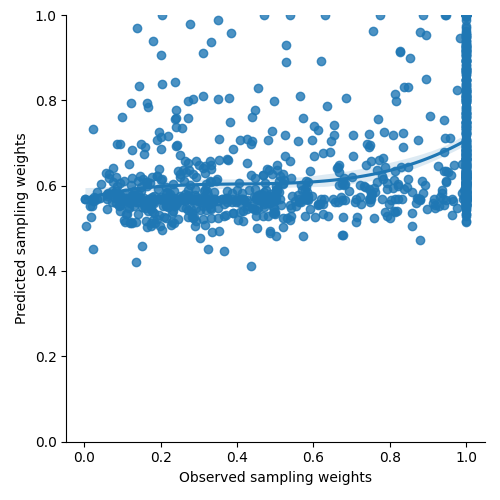}
\end{center}
\caption{Fitting the weights $w$}
\label{fig:w_fit_NN}
\end{figure}

Figure~\ref{fig:w_fit_NN} plots the fitted weights. As already mentioned,
the model for weights has low explanatory power. Many contracts cover the
entire year, as can be seen from the vertical cluster on the right. The
covariates do not help much in predicting how long the car is insured within
a given year, as it depends on decisions to buy, sell or exchange a vehicle
that result in starting or terminating coverage within a given calendar year.

In a linear model, we could use partial $R^{2}$'s to evaluate the
contribution of various covariates to explaining the left-hand side
variable. In a neural network, a natural alternative is to train the model
again while omitting one covariate and to measure the additional loss. This
is a very partial indication, as the neural network interacts different
groups of variables in potentially complex ways. Still, it is a reasonable
starting point.

Our application has eight groups of variables; accordingly, we run eight
neural networks, each of which omits one of these groups. Figure~\ref%
{fig:contribs_features_NN} plots the results. The dashed vertical line
represents the test loss of the complete neural network, which is denoted
\textquotedblleft None\textquotedblright . A larger positive number denotes
that this group of variables contributes more to the quality of the fit
(that is, reduces the test loss more).

\begin{figure}[tbp]
\begin{center}
\includegraphics{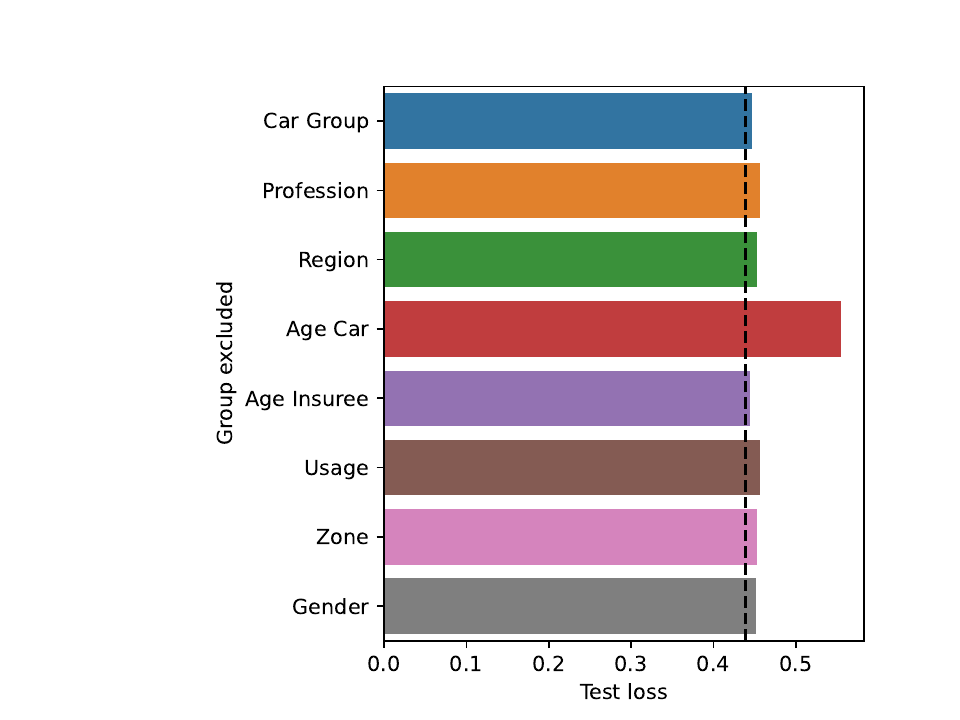}
\end{center}
\caption{Omitting groups of variables}
\label{fig:contribs_features_NN}
\end{figure}

{Figure~\ref{fig:contribs_features_NN} shows clearly that the age of the car
dominates the fit. Five other groups of variables contribute (taken by
themselves) to the fit. In decreasing order of importance, they are: the
(work/leisure) usage of the car, the profession, the rural/urban zone, the
age of the insuree, and her gender.} Looking more closely at the models for $%
c$ and $r$ shows that drivers of older cars are less likely to buy
comprehensive coverage---again, a common finding.

\subsection{Estimated covariance and correlation}\label{sub:covcorr:results}

We use the hyperoptimized models of Section~\ref{sub:hyperopt} to get ``raw'
and ``cross-fitted'' estimates for the covariances and correlation
functions. The raw estimates are simply the values predicted
over the whole sample. To obtain the cross-fitted estimates, we split the
sample randomly into five subsets; we predict the covariances and
correlations over a subset using the hyperoptimized neural network trained
over the other four subsets only. We will need the cross-fitted estimates
when we move to testing in Section~\ref{sub:intersection}. 

Table~\ref{tab:estimatesNN} gives the results. Only 40.4\% of the predicted
raw covariances (and correlations) are positive.  The
cross-fitted estimates are very similar to the raw estimates, if somewhat
more dispersed. 

\begin{table}[th]
\centering%
\begin{tabular}{l|ccc|ccc}
Name of variable & \multicolumn{3}{|c}{Raw estimates} & 
\multicolumn{3}{|c}{Cross-fitted estimates} \\ \cline{2-7}
& {Mean} & Dispersion & Range & {Mean}
& Dispersion & Range \\ 
\midrule Covariance & {-0.000} & 0.004 & [-0.014, 0.014]
& {-0.001} & 0.003 & [-0.019 , 0.016] \\ 
Correlation & {-0.007} & 0.043 & [-0.098, 0.092] & 
{-0.013} & 0.033 & [-0.121, 0.106] \\ 
\end{tabular}%
\caption{Raw and cross-fitted neural network estimates for the covariance
and correlation}
\label{tab:estimatesNN}
\end{table}

\begin{figure}[tbp]
\caption{Densities of $\hat{C}(x)$ and $\hat{\protect \rho}(x)$ over the
sample (raw neural network estimates)}
\label{fig:CovCorrPlotNN}%
\begin{subfigure}{\textwidth}
        \begin{center}
\includegraphics[scale=0.7]{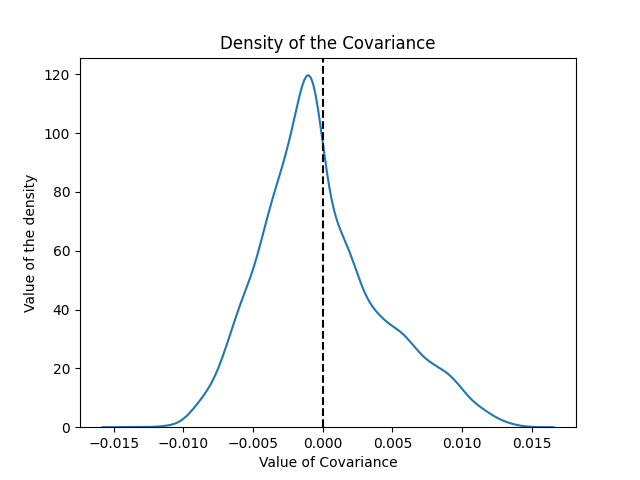}
\caption{Density of $\hat{C}(x)$}
    \end{center}
    \end{subfigure}
\par
\begin{subfigure}{\textwidth}
    \begin{center}
        \includegraphics[scale=0.7]{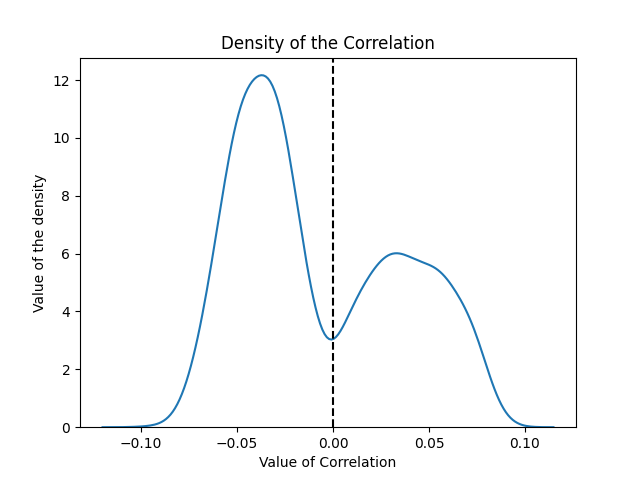}
        \caption{Density of $\hat{\rho}(x)$}
\end{center}
    \end{subfigure} 
\end{figure}

Figure~\ref{fig:CovCorrPlotNN} plots the density of our estimated covariance
function $\hat{C}(x)$ and correlation function $\hat{\rho}\left( x\right) $
over the sample.

Essentially all the mass of the distribution of the covariance is situated
in the interval $[-0.01,0.01]$. These seem like small values, but they are
not easily interpretable. We next consider the correlation function. It is
negatively skewed; more than 60\% of the estimates are negative. The
correlations are small, however: 99\% of the mass is in the $[-0.2,0.2]$
interval.

\begin{figure}[tbp]
\begin{center}
\includegraphics[scale=0.65]{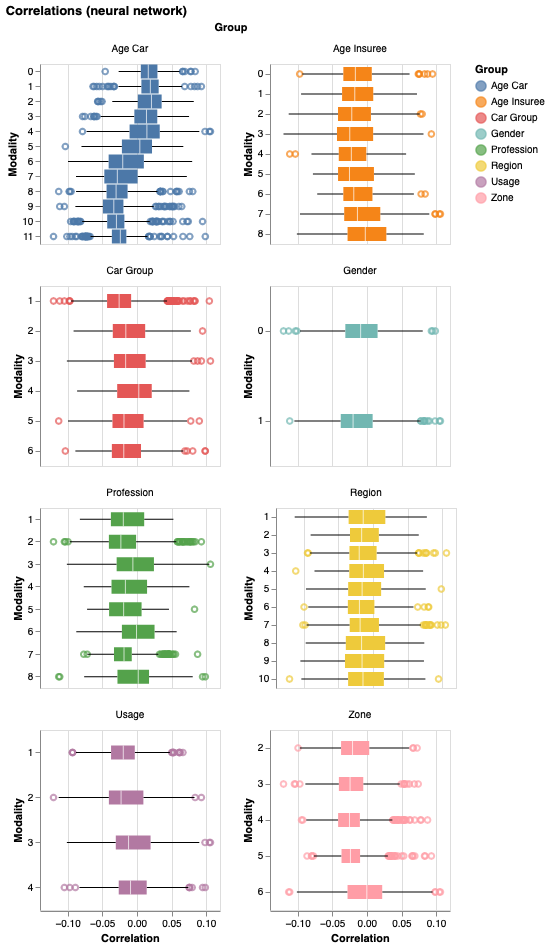}
\end{center}
\caption{Correlation $\hat{\protect \rho}(x)$ for different subgroups (raw
neural network estimates)}
\label{fig:boxesNN}
\end{figure}

It would be tempting to interpret the bimodal shape of the estimated density
of $\hat{\rho}$ as a mixture of the densities for the two genders. We
already know from Figure~\ref{fig:contribs_features_NN} that gender has low
explanatory power, however. Figure~\ref{fig:boxesNN} has boxplots of the
distributions of $\hat{\rho}$ when a variable takes one particular value.
The top-left panel, for instance, shows that the estimated correlation tends
to decrease with the age of the car. Men seem to have a lower correlation of
risk and coverage than women, rural drivers a lower correlation than urban
drivers.

\section{Nuisance parameters and double-debiasing}

\label{sec:covcorr} Our ultimate goal is to test the sign of the covariance
function $C(x)$ and the correlation function $\rho(x)$. These are easily
estimated by plugging in the probabilities $\hat{p}_{jk}(x)$ predicted by
the neural network. Still, these probability estimates have a relatively
slow rate of convergence, which is likely to contaminate inference. This is
a common issue with machine learning methods: they yield predictors that
must be treated as nuisance parameters in later stages of statistical
procedures. To remedy this problem, \citet{chernozhukov_doubledebiased_2018}
developed a double-debiasing method that extends the idea of Neyman
orthogonalization to a broad range of models and estimation methods.

To give the intuition behind double-debiasing, consider the vector of
probabilities $\eta \equiv (p_{00},p_{01}, p_{10}, p_{11})$. We denote $%
\eta_0 $ its true value. Suppose that we want to estimate some parameter
vector $\beta_0$ that satisfies a set of conditions $M(\beta_0,\eta_0)=0$.
In our application, this $\beta_0$ will be the average covariance or
correlation for a given subset of observations---for instance, for all young
men.

The neural network (or any other estimation procedure) gives us an estimate $%
\hat{\eta}$. A natural way to proceed would be to estimate $\beta_0$ by the
value $\hat{\beta}$ that minimizes some norm of $\hat{M}(\beta,\hat{\eta})$,
where the function $\hat{M}$ is the sample analog of $M$. Given an
appropriate set of assumptions, the standard Taylor expansion around the
true values $(\beta _{0},\eta _{0})$ gives 
\begin{equation}
\nabla _{\beta }M(\beta _{0},\eta _{0})(\hat{\beta}-\beta _{0})\simeq -
M(\beta _{0},\eta _{0})-\nabla _{\eta }M(\beta _{0},\eta _{0})( \hat{\eta}%
-\eta _{0}).  \label{eq:grad}
\end{equation}%
The presence of the second term on the right-hand side is what makes the $%
\hat{\eta}$ estimates nuisance parameters: the estimation error $\hat{\eta}$
contaminates the asymptotic distribution of $\hat{\beta}$ if the gradient $%
\nabla _{\eta }M(\beta _{0},\eta _{0})$ is nonzero.

If the $\hat{\eta}$ estimates converge at the usual parametric rate, the
additional term only changes the usual sandwich formula. This clearly does
not apply here, as the neural network estimates converge more slowly, and
this invalidates standard inference over $\beta _{0}$. To get rid of this
nuisance effect, we need to change the estimating equation to an equation
for which the gradient of $M$ with respect to $\eta $ is zero at the true
values $(\beta _{0},\eta _{0})$. This can be done by projecting the
estimating equation on the orthogonal subspace to the gradient $\nabla
_{\eta }\hat{M}$. This is the idea that underlies Neyman orthogonalization
and its modern successor, double-debiasing.

We describe how double-debiasing applies to the covariance and the
correlation functions in Sections~\ref{sub:getcov} and~\ref{sub:getcorr},
respectively.

\subsection{Covariance}

\label{sub:getcov} Given our estimates $\hat{p}_{jk}$ of the probabilities
of the four alternatives, we can write the covariance as 
\begin{equation*}
\widehat{C}(x)=\frac{\hat{E}\left( w(c-\hat{p}(x))(r-\hat{q}(x))|X=x\right) 
}{\hat{E}(w|X=x)}\equiv \hat{E}_{w}\left( (c-\hat{p}(x))(r-\hat{q}%
(x))|X=x\right) ,
\end{equation*}%
where $\hat{E}$ (resp.\ $\hat{E}_{w}$) denotes an unweighted (resp.\
weighted) sample mean; $\hat{p}(x)\equiv \hat{p}_{10}(x)+\hat{p}_{11}(x)$
estimates $\Pr (c=1|X=x)$ and $\hat{q}(x)\equiv \hat{p}_{01}(x)+\hat{p}%
_{11}(x)$ estimates $\Pr (r=1|X=x)$. In these equations, the neural network
estimates $\hat{p}_{jk}(x)$ act as nuisance parameters $\hat{\eta}$.
However, it turned out that for the covariance coefficients, the presence of
such parameters does not interfere in the estimation procedure, so that
standard inference methods apply.

To see this, consider estimating the sample-weighted average covariance $%
\beta_g$ for a group of observations $i\in g$. We will use weighted
non-linear least squares. This corresponds to the estimating equation 
\begin{equation*}
M(\hat{\beta}_g,\hat{\eta})=\sum_{i \in g}w_i \left(\widehat{C}(x_{i})-\hat{%
\beta}_g\right)=0.
\end{equation*}%
The gradient of $M$ with respect to $\eta$ is a weighted sum of gradients of 
$C(x)$ with respect to $\eta $. Now consider, for some $x$ and some $j,k=0,1$%
, 
\begin{equation*}
\frac{\partial C(x)}{\partial p_{jk}(x)}=-\frac{\partial p(x)}{\partial
p_{jk}(x)}E_{w}\left( (r-q(x))|X=x\right) -\frac{\partial q}{\partial p_{jk}}%
E_{w}\left( (c-p(x))|X=x\right) .
\end{equation*}%
By definition, at the true values $\eta _{0}$ we have 
\begin{equation*}
E_{w}\left( (r-q(x))|X=x\right) =E_{w}\left( (c-p(x))|X=x\right) =0.
\end{equation*}%
Since the partial derivatives of $p$ and $q$ with respect to $p_{jk}$ are
either 0 or 1, the gradient of $M$ with respect to $p_{jk}$ is zero at the
true values and the estimates of $\hat{\eta}$ do not affect the asymptotic
distribution of $\hat{\beta}$. To put it differently: the covariance is
doubly robust and does not need to double-debiased.

\subsection{Correlation}

\label{sub:getcorr} We now turn to the correlation function $\rho (x)$. A
naive estimate of the correlation function would be 
\begin{equation*}
\hat{\rho}(x)=\frac{\widehat{C}(x)}{\sqrt{\hat{p}(x)(1-\hat{p}(x))}\sqrt{%
\hat{q}(x)(1-\hat{q}(x))}}.
\end{equation*}%
The gradient of $\hat{\rho}$ with respect to $\hat{\eta}$ now involves terms
like 
\begin{equation*}
\frac{\partial \sqrt{\hat{p}(1-\hat{p})}}{\partial \hat{p}_{jk}}
\end{equation*}%
which are clearly not zero at the true values. Therefore the presence of
nuisance parameters does interfere in the estimation procedure and thus, can
invalidate inference unless we use double-debiasing.

Proceeding as with the covariance function, let our estimation equation be 
\begin{equation*}
M(\hat{\beta}_{g},\hat{\eta})=\sum_{i\in g}w_{i}\left( \hat{\rho}(x_{i})-%
\hat{\beta}_{g}\right) =0.
\end{equation*}%
Again, the gradient of $M$ with respect to $\eta $ is a weighted sum of
gradients of $\rho (x)$ with respect to $\eta (x)$. We already know that we
the derivatives of $C(x)$ with respect to $\eta (x)$ are zero. This leaves
us with 
\begin{equation*}
\frac{\partial \rho (x)}{\partial p_{jk}(x)}=\rho (x)\times \left( \frac{%
p(x)-1/2}{p(x)(1-p(x))}\frac{\partial p(x)}{\partial p_{jk}(x)}+\frac{%
q(x)-1/2}{q(x)(1-q(x))}\frac{\partial q(x)}{\partial p_{jk}(x)}\right) .
\end{equation*}%
To apply double-debiasing, we need to project our estimating equation on the
orthogonal space to this gradient. This can be done quite simply by running
a weighted regression of $\hat{\rho}(x_{i})$ on the variables that
correspond to $(j,k)=(0,1),(1,0)$, and $(1,1)$. In fact, the last one is the
sum of the previous two, so that we only need to use the following two
regressors\footnote{%
Using $p=p_{10}+p_{11}$ and $q=p_{01}+p_{11}$.}: 
\begin{equation*}
\hat{\nabla}_{1i}\equiv \hat{\rho}_{i}\frac{\hat{q}_{i}-1/2}{\hat{q}%
_{i}\left( 1-\hat{q}_{i}\right) }\; \mbox{ and }\; \hat{\nabla}_{2i}\equiv 
\hat{\rho}_{i}\frac{\hat{p}_{i}-1/2}{\hat{p}_{i}\left( 1-\hat{p}_{i}\right) }%
.
\end{equation*}%
To obtain a double-debiased estimator of $\beta _{g}=E_{w}(\rho (x_{i})|i\in
g)$, we can simply regress $\hat{\rho}_{i}$ on $\hat{\nabla}_{1i}$, $\hat{%
\nabla}_{2i}$, and a constant over the sample $i\in g$ with weights $w_{i}$.
The double-debiased estimator is the coefficient of the constant; we will
denote it $\tilde{\rho}_{g}$ from now on on.

\section{Testing the Positive Correlation Property}

\label{sec:PCP} We implement two ways of testing the positive correlation
property: an intersection test and a sorted group approach, in Sections \ref%
{sub:intersection} and~\ref{sub:sortedgroups} respectively. 
Both methods require that we use cross-fitted predictors of the covariance
and correlation; we use the estimates described in Section~\ref%
{sub:estimates} for this purpose.

\subsection{The intersection test}\label{sub:intersection} 
It is clearly not feasible to test that the
covariance (or correlation) is positive for all possible values of all
covariates. However, results in \citet{SemenovaCherno:EJ2021} show that we
can use standard inference for their mean values over subgroups of
observations.

More precisely, let $T$ denote either covariance or correlation, and
consider testing the following hypothesis: 
\begin{equation}
C(\bar{h})\equiv E(T(X)|h(X)=\bar{h})\; \; \forall \bar{h}\text{ in }B,
\label{eq:cov:h}
\end{equation}%
where $h$ is a function whose values lie in a low-dimensional space, and $B$
is a subset of its range. We could for instance test that the positive
correlation property holds on average over all women in rural areas who
drive cars that are more than 5 year old.

\citet{SemenovaCherno:EJ2021} derive assumptions under which one can run a
sieve regression 
\begin{equation*}
\hat{T}(x_{i})=p(h(x_{i}))\beta +u_{i}
\end{equation*}%
and use its fitted values $p(h(x_{i}))\hat{\beta}$, instead of the neural
network estimate of $T(x_{i})$, to test the multiple hypothesis %
\eqref{eq:cov:h}. Note that $p$ applies to the values of $h$; in the example
of the previous paragraph, $p$ could only be a function of gender, whether
the insuree lives in a rural area, and whether the car is more than 5 years
old.

The statistic $T$ must be double-debiased if needed (that is, we use $\hat{C}
$ and $\tilde{\rho}$); it must converge faster than $n^{-1/4}$, which holds
under reasonable assumptions on the neural network; and the sieve basis $p$
must expand at the appropriate rate, which implicitly limits the
dimensionality of the function $h$.

The testing procedure simplifies further if we focus on groups of
observations. To be more precise, suppose that we define disjoint groups $%
g_{1},\ldots ,g_{L}$ by the values of some of our covariates. The union of
the groups could be the whole set of observations, but that is not
necessary. For instance, $g_{1}$ could have all young men and $g_{2}$ all
women who have an old car. Our goal is to test the null hypothesis that 
\begin{equation*}
E(T(X)|X\in g_{l})\geq 0\; \text{ for }\;l=1,\ldots ,L
\end{equation*}%
where $T$ is either the covariance or the correlation. In the notation of
the previous paragraphs, this amounts to using indicators of the groups as
the basis functions $p\circ h$ in the sieve regression; the model is
saturated and the sieve basis trivially satisfies the conditions in %
\citet{SemenovaCherno:EJ2021}.

Then for each of our groups $g_{l}$, we compute the sample-weighted average
predicted covariance $\hat{C}_{l}$. We apply the double-debiasing procedure
described in the previous subsection to obtain $\tilde{\rho}_{l}$, the
double-debiased, sample-weighted average correlation within this group. We
compute the standard error $\hat{\sigma}_{l}$ of the estimators $\hat{T}_{l}=%
\tilde{\rho}_{l}$ or $\hat{C}_{l}$ by the usual formula.

If $L=1$ (for instance, we only want to test the positive correlation
property for young men), then we are done: we reject the null hypothesis at
the 5\% level if $\hat{T}_{l}+1.64\hat{\sigma}_{l}<0$. If $L>1$, we want to
test that $\min_{l=1,\ldots ,L}E(T(X)|X\in g_{l})>0$. This is an
intersection test; we use the procedure described in \cite%
{chernozhukov_intersection_2013}:

\begin{enumerate}
\item we draw a large number of values $(\xi _{r})_{r=1,\ldots ,R}$ from $%
N(0,I_{L})$;

\item for each $r$, we define $\bar{v}_{r}=\max_{l=1,\ldots ,L}\xi _{rl}$;
we let $k_{0}$ be the $\gamma _{n}$-quantile of the $\bar{v}_{r}$, with $%
\gamma _{n}=1-0.1/\log n$;

\item we let $\hat{L}$ be the set of values of $l$ such that 
\begin{equation*}
\hat{T}_{l}\leq \min_{m=1,\ldots ,L}\left( \hat{T}_{m}+k_{0}\hat{\sigma}%
_{m}\right) +2k_{0}\hat{\sigma}_{l};
\end{equation*}

\item finally, we let $k$ be the $(1-\alpha )$-quantile of the values 
\begin{equation*}
\hat{v}_{r}=\max_{l\in \hat{L}}\xi _{rl}
\end{equation*}%
and we reject the hypothesis if $\inf_{l\in \hat{L}}\left( \hat{T}_{l}+k\hat{%
\sigma}_{l}\right) <0$.
\end{enumerate}

In Sections~\ref{ssub:cov} and~\ref{ssub:corr}, we will implement the
intersection test for the covariance and the correlation coefficient,
respectively.{\ To define the groups $g_{l}$, we use the age of the car,
which is by far the variable with the most predictive content.} We split it
into quartiles, and we also run tests for all of its modalities.

\subsubsection{The results of the intersection test for the covariance}\label{ssub:cov} 
 Table~\ref{tab:cvalsinter:covNN} gives the results of the
intersection tests for the variable \textquotedblleft car
age\textquotedblright \ and its split by quartiles (the results of the
intersection tests for other variables are provided in Table ~\ref%
{tab:cvalsinter:covNNother} in the Appendix)\footnote{%
The bounds of the confidence intervals are computed as $a=\min_{l}(\hat{C}%
_{l}+k\hat{\sigma}_{l})$ and $b=\max_{l}(\hat{C}_{l}-k\hat{\sigma}_{l})$:
these are the values such that we are at the margin of rejecting the
hypothesis that $\min_{l}C_{l}>a$ and the hypothesis that $\max_{l}C_{l}<b$.}%
.

\begin{table}[th]
\centering%
\begin{tabular}{l|ccccccc}
 Group & Modalities & Test level & $k_{0}$ & $k$
& Test statistic & PCP & Confidence interval \\ 
\midrule Car age & {4} & 0.01 & 2.71 & 2.17 & -0.0023 & 
rejected & [-0.0023,0.0011] \\ 
quartiles &  & 0.05 & 2.71 & 1.57 & -0.0023 & rejected
& [-0.0023,0.0012] \\ 
&  & 0.10 & 2.71 & 1.28 & -0.0023 & rejected & 
[-0.0023,0.0012] \\ \hline
Car age & 12 & 0.01 & 3.12 & 2.94 & -0.0026 & rejected
& [-0.0025,0.0011] \\ 
& & 0.05 & 3.12 & 2.41 & -0.0026 & rejected & 
[-0.0026,0.0012] \\ 
&  & 0.10 & 3.12 & 2.08 & -0.0027 & rejected & 
[-0.0026,0.0013] \\ 
\end{tabular}%
\caption{PCP test using the covariances for the ``age of the car'' variable
variable: neural network estimates}
\label{tab:cvalsinter:covNN}
\end{table}

Note that $k$ (the number of standard errors used in the last step of the
intersection test) varies noticeably. For the 5\% test, it varies from $1.62$
(that is, the standard value) to $2.54$, which is larger than what a naive
normal approximation would suggest. The test statistics show that for any of
the partitions into groups, we can reject the hypothesis that the
covariances are all positive at any reasonable level. On the other hand, we
can also reject that any of them is smaller (i.e., less negative) than, say, 
$-0.01$; this can be seen in the \textquotedblleft confidence
interval\textquotedblright \ column.


\subsubsection{The results of the intersection test for the correlation}
\label{ssub:corr}
 Let us turn to the correlation function. The results in
Table~\ref{tab:cvalsinter:rhoNN} are very similar to those for the
covariance: once again, the positive correlation property is rejected for
all partitions of the sample (the test results for other variables are shown
in Table~\ref{tab:cvalsinter:rhoNNother} in the Appendix).

\begin{table}[th]
\centering%
\begin{tabular}{l|ccccccc}
 Group & {Modalities} & Test level & $k_{0}$ & $k$
& Test statistic & PCP & Confidence interval \\ 
\midrule Car age & {4} & 0.01 & 2.71 & 2.61 & -0.0034 & 
rejected & [-0.0034,0.0003] \\ 
quartiles & {} & 0.05 & 2.71 & 2.10 & -0.0036 & rejected
& [-0.0035,0.0004] \\ 
& {} & 0.10 & 2.71 & 1.77 & -0.0037 & rejected & 
[-0.0036,0.0005] \\ \hline
Car age & {12} & 0.01 & 3.12 & 2.99 & -0.0035 & rejected
& [-0.0035,0.0012] \\ 
& {} & 0.05 & 3.12 & 2.52 & -0.0038 & rejected & 
[-0.0037,0.0016] \\ 
& {} & 0.10 & 3.12 & 2.20 & -0.0040 & rejected & 
[-0.0038,0.0018] \\ 
\bottomrule &  &  &  &  &  &  & 
\end{tabular}%
\caption{PCP test using the correlation coefficient for the ``car age of the
car'' variable: neural network estimates }
\label{tab:cvalsinter:rhoNN}
\end{table}

Moreover, the confidence intervals confine the (doubly-debiased) correlation
to a narrow interval just below zero. This is not at all what Figure~\ref%
{fig:CovCorrPlotNN} (which plots the density of $\hat{\rho}$, not that of
the double-debiased version $\tilde{\rho}$) would have suggested; it is a
good illustration of the fact that raw predictions from neural networks are
not free from bias and noise.

\subsection{The sorted groups approach}\label{sub:sortedgroups} 
A recent contribution by \cite%
{chernodemirerduflofv23} adopts a different approach to recover estimators
of group averages that have standard asymptotics. While they define it in
the more complex framework of conditional average treatment effects, their
method a fortiori applies in our setting. The underlying idea is to split
the sample into a main sample and an auxiliary subsample. A machine learning
model is trained and optimized on the auxiliary subsample to predict the
statistics of interest (here, the probabilities of the four alternatives).
The predictors are applied to the observations in the main sample, which are
then allocated into groups sorted by the value of, in our case, the
predicted covariance or correlation. \cite{chernodemirerduflofv23} show that
regressing the outcomes $y_{jk}$ observed in the main sample on group
indicators gives estimators of the average probabilities that have standard
asymptotics and can therefore be used to construct standard tests.

\subsubsection{The results of the sorted groups approach for the covariance}\label{sub:sortedcov} 
As recommended by \cite{chernodemirerduflofv23}, we
use several random splits into main and auxiliary samples, and we report the
median test statistics and $p$-values. We train and select a neural network
on the auxiliary subsample exactly as explained in Section~\ref{sec:nn}.
Once we have predicted probabilities $\hat{p}_{jk}$ on the main sample we
use them to compute the covariance for each observation using the formul\ae \ %
in Section~\ref{sub:getcov}. We then sort the observations according to the
predicted covariance, and we define four groups $q=1,2,3,4$, splitting at
the quartiles. In each group, we regress $y_{jk}$ on the group indicators to
obtain new predicted group-average probabilities $\bar{p}_{jk}(q)$. The
results in~\cite{chernodemirerduflofv23} imply that these $\bar{p}$
statistics have standard asymptotics. This allows us to test the PCP on each
group $q$, and to define confidence intervals for the $q$-group covariances.

The results are reported in Table~\ref{tab:sortedgroupscov}, where the
medians are computed over 100 random splits. While the PCP is not rejected
at the 5\% level, the confidence interval for the average covariance on the
lowest-covariance quartile is very narrow. This is consistent with the
results in Tables~\ref{tab:cvalsinter:covNN} and~\ref{tab:cvalsinter:rhoNN}.

\medskip

\begin{table}[th]
\centering%
\begin{tabular}{lccccc}
Statistic & {Covariance} & Estimated & 95\%
confidence & Test statistic & $p$-value \\ 
& {in quartile 1} & standard error & interval &  &  \\ 
\toprule
Median over 101 splits & {$-0.0030$} & $0.0024$ & $%
[-0.0076,0.0016]$ & $-1.28$ & $0.10$ \\ 
\end{tabular}%
\caption{Testing for a positive covariance on the first covariance quartile}
\label{tab:sortedgroupscov}
\end{table}

\subsubsection{The results of the sorted groups approach for the correlation}
\label{sub:sortedcorr} 
We proceed in exactly the same way for the
correlation. Remarkably, the main/auxiliary method used in \cite%
{chernodemirerduflofv23} allows us to circumvent double-debiasing
altogether. As Table~\ref{tab:sortedgroupscorr} shows, we come close to
rejecting the PCP for the lowest-correlation quartile.

\medskip

\begin{table}[th]
\centering%
\begin{tabular}{lccccc}
Statistic & {Correlation} & Estimated & 95\%
confidence & Test statistic & $p$-value \\ 
& {in quartile 1} & standard error & interval &  &  \\ 
\toprule
Median over 101 splits & {$-0.0304$} & $0.0207$ & $%
[-0.0672,0.0083]$ & $-1.52$ & $0.06$ \\ 
\end{tabular}%
\caption{Testing for a positive correlation on the first correlation quartile
}
\label{tab:sortedgroupscorr}
\end{table}

The confidence interval shows once more that only very small values of the
correlation are consistent with the data.

\section{Deep learning vs tree-based methods}
\label{sec:RFandGBT}
 For the sake of comparison, in this section we use another popular machine approach: ensemble learning
applied to decision trees. To train the ensemble, we consider two
alternative learners: bagging and boosting. Bagging applied to feature
selection using decision trees is referred to as \textquotedblleft random
forests\textquotedblright , while the boosting method yields
gradient-boosted trees. Using these two methods, we estimate the covariance
and the correlation functions, we double-debias the correlation, and we test
the positive correlation property using the intersection test.

\subsection{Model selection}
In the machine learning setting, predictions from all types of models can be
averaged. This approach has become especially popular with models based on
decision trees --- ensemble learners that average over a large numbers of
shallow decision trees that are trained over random subsamples and features.%

\begin{figure}[tbp]
\begin{center}
\includegraphics{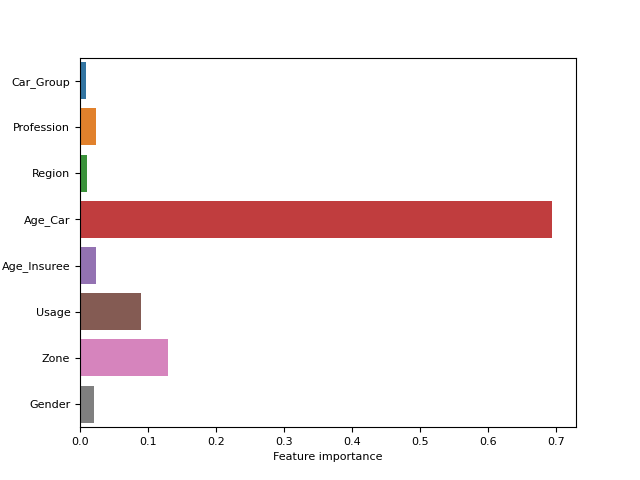}
\end{center}
\caption{Feature importances for the best random forest}
\label{fig:feature_importances_random_forest}
\end{figure}

We use $N=500$ ensemble members and 5-fold cross validation with the
weighted entropy criterion. We do a grid search over three hyperparameter
values: the maximum depth of each decision tree; the minimum number of
observations in each leaf; and the maximum number of features that are
randomly selected before each split (for the random forest) or the learning
rate (for the gradient-boosted tree). To optimize in hyperparameter space,
we split the data into training and testing sets, representing 80\% and 20\%
of the data respectively; we select the model that gives the best fit on a
test sample. Both procedures report ``feature importance'' scores; we plot
them in Figures~\ref{fig:feature_importances_random_forest} and~\ref%
{fig:feature_importances_gradient_boosted_tree}.

\begin{figure}[tbp]
\begin{center}
\includegraphics{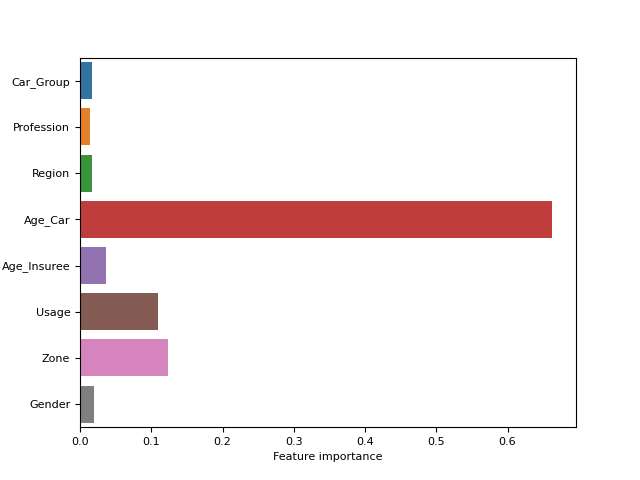}
\end{center}
\caption{Feature importances for the best gradient boosting method}
\label{fig:feature_importances_gradient_boosted_tree}
\end{figure}

The best random forest has a maximum depth of 5; a minimum leaf size of 10;
and randomization over a maximum of 5 features. Figure~\ref%
{fig:feature_importances_random_forest} shows that the age of the car again
is the most important explanatory variable by far. The best gradient-boosted
tree has a maximum depth of 4; a minimum leaf size of 20; and a learning
rate of $0.1$. Figure~\ref{fig:feature_importances_gradient_boosted_tree}
shows that the importance of the various features for the gradient boosting
is very similar to that for our selected random forest.

When computed over the whole sample, the weighted entropy losses of the
three methods are $0.386$ for the neural network, $0.392$ for the
gradient-boosted tree, and $0.402$ for the random forest. We were surprised
that the tree-based learners do not work better on this tabular data. These
differences are not very large, however, and they may be quite specific to
our dataset.

\subsection{Comparison results for the covariance and correlation functions}
Table~\ref{tab:estimatesTrees} reports our estimates of the covariances and
correlation coefficients produced by random forest and gradient boosting
methods.

\begin{table}[th]
\centering%
\begin{tabular}{l|ccc|ccc}
Name of variable & \multicolumn{3}{|c}{Random Forest} & 
\multicolumn{3}{|c}{Gradient Boosting} \\ \cline{2-7}
& {Mean} & Dispersion & Range & {Mean}
& Dispersion & Range \\ 
\midrule Covariance & {-0.002} & 0.004 & [-0.028, 0.012]
& {-0.003} & 0.009 & {[-0.066, 0.069]}
\\ 
Correlation & {-0.018} & 0.035 & [-0.198, 0.107] & 
{-0.032} & 0.083 & {[-0.420, 0.503]}
\\ 
\end{tabular}%
\caption{Raw random-forest and gradient-boosting estimates for the
covariance and correlation}
\label{tab:estimatesTrees}
\end{table}

\begin{figure}[tbp]
\begin{center}
\includegraphics{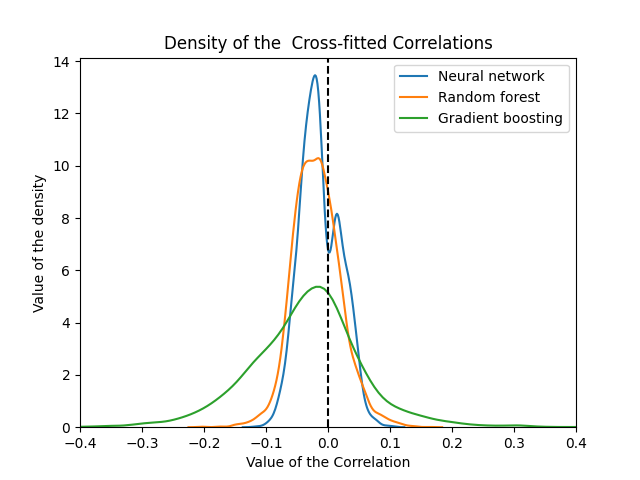}
\end{center}
\caption{Density of $\hat{\protect \rho}(x)$ for the three machine learners}
\label{fig:correl_density_plots}
\end{figure}

Figure~\ref{fig:correl_density_plots} plots the density of the predicted
correlations $\hat{\rho}_{i}$ over the sample under all three of our machine
learning methods: deep learning, random forest, and gradient boosting. The
range of values of the correlation coefficient is similar for all three
methods: essentially all mass belongs to the $[-0.2,0.2]$ interval. While
the density of $\rho (x)$ obtained from the neural network is bimodal, with
two asymmetric peaks in the negative and positive ranges, the densities
produced by the ensembles are unimodal. Another important difference is that
the gradient-boosted tree produces a much broader range of variation than
the other two methods on this data.

\begin{figure}[tbp]
\begin{center}
\includegraphics[scale=0.65]{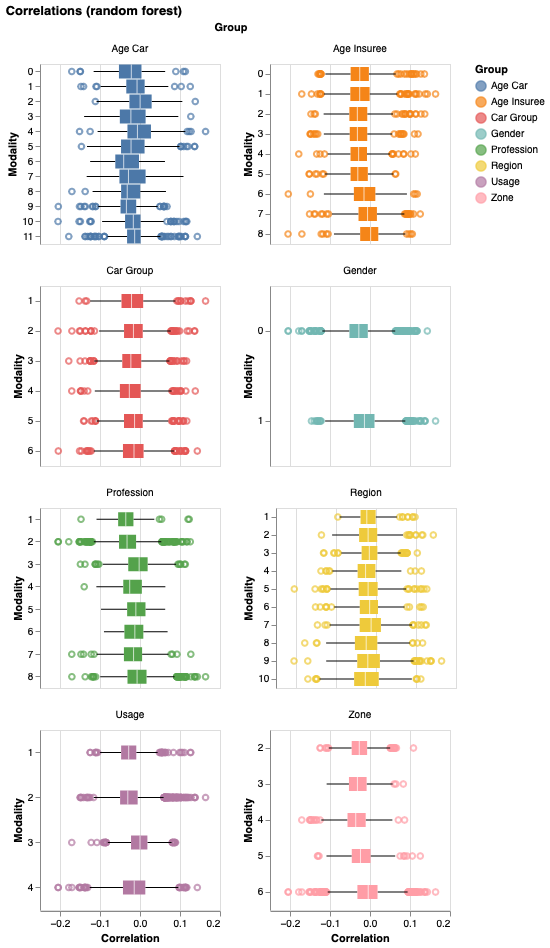}
\end{center}
\caption{Correlation $\protect \rho (x)$ obtained from the random forest for
different groups}
\label{fig:boxes_correl_RF}
\end{figure}

\begin{figure}[tbp]
\begin{center}
\includegraphics[scale=0.65]{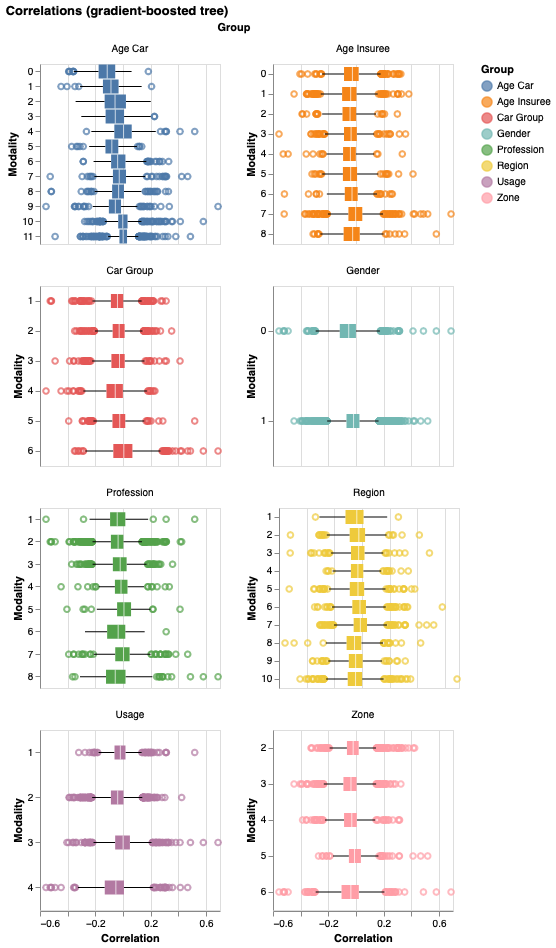}
\end{center}
\caption{Correlation $\protect \rho (x)$ obtained from gradient boosting for
different groups}
\label{fig:boxes_correl_GBT}
\end{figure}

Figure~\ref{fig:boxes_correl_RF} (resp.\ Figure~\ref{fig:boxes_correl_GBT})
shows the correlation for different groups produced by the random forest
method (resp.\ the gradient boosting method). These plots show some marked
differences with the equivalent plot for the neural network (Figure~\ref%
{fig:boxesNN}), most notably for the important \textquotedblleft Age of
car\textquotedblright \ variable.

\subsection{The intersection test}

We repeated the intersection tests of the positive correlation property with
our random forest and our gradient-boosted tree. We focus on the correlation
to save space. Table~\ref{tab:RFintersectiontestCOR} reports the results for
the random forest method, and Table~\ref{tab:GBTintersectiontestCOR} gives
them for the gradient boosting method (the results for other variables are
provided in the Appendix).

\begin{table}[th]
\centering%
\begin{tabular}{l|ccccccc}
    Group & {Modalities} & Test level & $k_{0}$ & $k$
   & Test statistic & PCP & Confidence interval \\ 
\midrule Car age & {4} & 0.01 & 2.80 & 2.37 & -0.0033 & 
rejected & [-0.0033,-0.0003] \\ 
quartiles & {} & 0.05 & 2.80 & 1.63 & -0.0034 & rejected
& [-0.0034,-0.0002] \\ 
& {} & 0.10 & 2.80 & 1.23 & -0.0035 & rejected & 
[-0.0034,-0.0001] \\ \hline
Car age & {12} & 0.01 & 3.18 & 2.18 & -0.0074 & rejected
& [-0.0073,0.0044] \\ 
& {} & 0.05 & 3.18 & 1.68 & -0.0077 & rejected & 
[-0.0075,0.0045] \\ 
& {} & 0.10 & 3.18 & 1.21 & -0.0080 & rejected & 
[-0.0077,0.0045] \\ 
\end{tabular}%
\caption{PCP test using the correlation coefficient for the ``car age''
variable: random forest estimation}
\label{tab:RFintersectiontestCOR}
\end{table}

\begin{table}[th]
\centering%
\begin{tabular}{l|ccccccc}
 Group & {Modalities} & Test level & $k_{0}$ & $k$
& Test statistic & PCP & Confidence interval \\ 
\midrule Car age & {4} & 0.01 & 2.80 & 2.67 & -0.0159 & 
rejected & [-0.0158,-0.0086] \\ 
quartiles & {} & 0.05 & 2.80 & 1.94 & -0.0160 & rejected
& [-0.0159,-0.0085] \\ 
& {} & 0.10 & 2.80 & 1.64 & -0.0160 & rejected & 
[-0.0160,-0.0085] \\ \hline
Car age & {12} & 0.01 & 3.18 & 2.75 & -0.0262 & rejected
& [-0.0255,0.0183] \\ 
& {} & 0.05 & 3.18 & 2.13 & -0.0278 & rejected & 
[-0.0272,0.0189] \\ 
& {} & 0.10 & 3.18 & 1.85 & -0.0286 & rejected & 
[-0.0278,0.0191] \\
\end{tabular}%
\caption{PCP test using the correlation coefficient for the ``car age''
variable: gradient-boosted estimation}
\label{tab:GBTintersectiontestCOR}
\end{table}

Since the lower bounds are negative for all methods, we can reject the
hypothesis that the covariances are all positive. The lower bounds given by
the random forest and especially the gradient-boosted tree are lower than
with the neural network; still, even with the gradient-boosted tree the
lower bound is not economically significant from zero. We can safely
conclude that the three machine learning methods produce qualitatively
similar implications. 

\subsection{The group-averaged correlations}

Finally, Tables~\ref{tab:CarAgeQuartiles} and~\ref{tab:CarAge}
report the group-averaged correlations produced by the three machine
learners at two levels of grouping on the \textquotedblleft Age of the
car\textquotedblright \ variable: with respectively four and twelve
modalities\footnote{%
For completeness, the Appendix compares the group-averaged correlations
produced by the three methods for the other variables.}. The great majority
of the correlations are negative, and they tend to shrink after
double-debiasing (indicated by ``DD'').

\begin{table}[th]
\centering%
\begin{tabular}{r|cc|cc|cc}
& \multicolumn{2}{c}{Neural network} & \multicolumn{2}{c}{Random forest} & 
\multicolumn{2}{c}{Gradient-boosted tree} \\ 
\toprule Modality & {Raw} & DD & {Raw}
& DD & {Raw} & DD \\ 
\midrule1 & {0.0152} & {$\underset{%
(0.0003)}{0.0009}$} & -0.0106 & $\underset{(0.0002)}{-0.0014}$ & -0.0640 & $%
\underset{(0.0009)}{-0.0124}$ \\ 
2 & {-0.0139} & {$\underset{(0.0002)}{%
-0.0041}$} & -0.0236 & $\underset{(0.0002)}{0.0002}$ & -0.0399 & $\underset{%
(0.0005)}{-0.0078}$ \\ 
3 & {-0.0300} & {$\underset{(0.0003)}{%
-0.0042}$} & -0.0232 & $\underset{(0.0002)}{-0.0037}$ & -0.0235 & $\underset{%
(0.0001)}{-0.0162}$ \\ 
4 & {-0.0263} & {$\underset{(0.0004)}{%
-0.0028}$} & -0.0148 & $\underset{(0.0001)}{-0.0025}$ & 0.0019 & $\underset{%
(0.0002)}{-0.0080}$
\end{tabular}%
\caption{Group-averaged correlations: car age quartiles}
\label{tab:CarAgeQuartiles}
\end{table}

\begin{table}[th]
\centering%
\begin{tabular}{r|cc|cc|cc}
& \multicolumn{2}{c}{Neural network} & \multicolumn{2}{c}{Random forest} & 
\multicolumn{2}{|c}{Gradient-boosted tree} \\ 
\toprule Modality & {Raw} & DD & {Raw}
& DD & {Raw} & DD \\ 
\midrule0 & {0.0180} & {$\underset{%
(0.0007)}{0.0033}$} & -0.0243 & $\underset{(0.0007)}{-0.0089}$ & -0.1206 & $%
\underset{(0.0027)}{-0.0337}$ \\ 
1 & {0.0173} & {$\underset{(0.0006)}{%
0.0018}$} & -0.0106 & $\underset{(0.0003)}{-0.0022}$ & -0.0911 & $\underset{%
(0.0020)}{-0.0200}$ \\ 
2 & {0.0180} & {$\underset{(0.0007)}{%
0.0013}$} & 0.0013 & $\underset{(0.0002)}{0.0046}$ & -0.0549 & $\underset{%
(0.0020)}{-0.0085}$ \\ 
3 & {0.0116} & {$\underset{(0.0005)}{%
-0.0003}$} & -0.0169 & $\underset{(0.0007)}{0.0009}$ & -0.0561 & $\underset{%
(0.0019)}{-0.0102}$ \\ 
4 & {0.0110} & {$\underset{(0.0006)}{%
0.0011}$} & -0.0012 & $\underset{(0.0002)}{0.0051}$ & 0.0075 & $\underset{%
(0.0009)}{0.0205}$ \\ 
5 & {-0.0012} & {$\underset{(0.0003)}{%
-0.0029}$} & -0.0156 & $\underset{(0.0004)}{0.0032}$ & -0.0793 & $\underset{%
(0.0019)}{-0.0152}$ \\ 
6 & {-0.0155} & {$\underset{(0.0004)}{%
-0.0034}$} & -0.0331 & $\underset{(0.0004)}{0.0003}$ & -0.0272 & $\underset{%
(0.0007)}{0.0027}$ \\ 
7 & {-0.0226} & {$\underset{(0.0005)}{%
-0.0045}$} & -0.0209 & $\underset{(0.0003)}{-0.0018}$ & -0.0199 & $\underset{%
(0.0005)}{-0.0096}$ \\ 
8 & {-0.0276} & {$\underset{(0.0005)}{%
-0.0051}$} & -0.0252 & $\underset{(0.0003)}{-0.0039}$ & -0.0342 & $\underset{%
(0.0006)}{-0.0167}$ \\ 
9 & {-0.0320} & {$\underset{(0.0005)}{%
-0.0033}$} & -0.0302 & $\underset{(0.0003)}{-0.0030}$ & -0.0548 & $\underset{%
(0.0010)}{-0.0139}$ \\ 
10 & {-0.0302} & {$\underset{(0.0005)}{%
-0.0037}$} & -0.0175 & $\underset{(0.0002)}{-0.0023}$ & 0.0034 & $\underset{%
(0.0002)}{-0.0078}$ \\ 
11 & {-0.0263} & {$\underset{(0.0004)}{%
-0.0028}$} & -0.0148 & $\underset{(0.0001)}{-0.0025}$ & 0.0019 & $\underset{%
(0.0002)}{-0.0080}$ \\
\end{tabular}%
\caption{Group-averaged correlations: car age}
\label{tab:CarAge}
\end{table}

\section*{Concluding Remarks}
Even with the very flexible methods used in our paper, this dataset shows no evidence for the positive correlation
property. In addition to this empirical finding, our paper contains a methodological
contribution. When we started this project, it was not clear to us that
deep learning methods could be applied fruitfully to a non-trivial testing
problem on such a (relatively) small dataset. With our sample of just $6,333$
observations, deep learning cannot go very deep:  neural networks with at
most two hidden layers and a small dropout rate work best. Still, they
deploy many more parameters than any econometric model.  Our neural
network in fact predict the  choice of coverage  much better than parametric procedures, and they also 
outperform them on  claim occurrence.  Double-debiasing,
or sorted groups, give us consistent and asymptotically normal estimates
that can be used in standard test procedures. Our overall conclusion is that
deep learning can provide both robust and useful conclusions even for
relatively small-scale applications.  Methods based on decision trees---random forests 
and gradient boosting---also perform well,  and give very similar results. In
further work, we plan to apply our three-step method to larger samples of
insurees, and to explore the value of reducing the number of covariates to
those that seem to have the largest effect.

\newpage

\bibliographystyle{econometrica}
\bibliography{insuranceNN}

\newpage

\section*{Appendix: additional results}

Tables~\ref{tab:cvalsinter:covNNother} and~\ref%
{tab:cvalsinter:rhoNNother} show the results of intersection tests for all
groups of variables except the age of the car (for which we reported the
results  in Tables~\ref{tab:cvalsinter:covNN} and~\ref{tab:cvalsinter:rhoNN}
in the main text), using our neural network estimates. Table~\ref%
{tab:cvalsinter:covNNother} focuses on covariances and Table~\ref%
{tab:cvalsinter:rhoNNother} on correlations. 

Tables~\ref{tab:RFintersectiontestCOV} and~\ref{tab:GBTintersectiontestCOV}
shows the results of the intersection test for group-averaged correlation
functions for the random forest and the gradient-boosted tree. Finally,
Tables~\ref{tab:CarGroup} to~\ref{tab:zone} give the values of the
group-averaged correlations for all modalities of other variables than the
age of the car, for the three machine-learning methods\footnote{%
Results for the covariances are available upon request.}.

\bigskip

\begin{table}[th]
\centering%
\begin{tabular}{l|ccccccc}
 Group & {Modalities} & Test level & $k_{0}$ & $k$
& Test statistic & PCP & Confidence interval \\ 
\midrule Car group & {6} & {0.01} & 
{2.96} & {2.85} & {
-0.0014} & {rejected} & {
[-0.0014,-0.0006]} \\ 
& {} & {0.05} & {2.96}
& {2.31} & {-0.0015} & 
{rejected} & {[-0.0014,-0.0006]} \\ 
& {} & {0.10} & {2.96}
& {2.00} & {-0.0015} & 
{rejected} & {[-0.0015,-0.0005]} \\ 
\hline
Insuree age & {9} & {0.01} & 
{3.11} & {3.07} & {
-0.0010} & {rejected} & {
[-0.0009,-0.0005]} \\ 
& {} & {0.05} & {3.11}
& {2.58} & {-0.0010} & 
{rejected} & {[-0.0010,-0.0005]} \\ 
& {} & {0.10} & {3.11}
& {2.27} & {-0.0011} & 
{rejected} & {[-0.0010,-0.0004]} \\ 
\hline
Gender & {2} & {0.01} & 
{2.60} & {2.61} & {
-0.0009} & {rejected} & {
[-0.0009,-0.0008]} \\ 
& {} & {0.05} & {2.60}
& {2.00} & {-0.0009} & 
{rejected} & {[-0.0009,-0.0008]} \\ 
& {} & {0.10} & {2.60}
& {1.65} & {-0.0010} & 
{rejected} & {[-0.0009,-0.0008]} \\ 
\hline
Zone & {5} & {0.01} & 
{2.89} & {2.80} & {
-0.0014} & {rejected} & {
[-0.0013,-0.0004]} \\ 
& {} & {0.05} & {2.89}
& {2.26} & {-0.0014} & 
{rejected} & {[-0.0014,-0.0004]} \\ 
& {} & {0.10} & {2.89}
& {1.99} & {-0.0014} & 
{rejected} & {[-0.0014,-0.0003]} \\ 
\hline
Usage & {4} & {0.01} & 
{2.71} & {2.61} & {
-0.0011} & {rejected} & {
[-0.0011,-0.0007]} \\ 
& {} & {0.05} & {2.71}
& {2.02} & {-0.0012} & 
{rejected} & {[-0.0012,-0.0006]} \\ 
& {} & {0.10} & {2.71}
& {1.70} & {-0.0012} & 
{rejected} & {[-0.0012,-0.0006]} \\ 
\hline
Profession & {8} & {0.01} & 
{3.03} & {2.97} & {
-0.0012} & {rejected} & {
[-0.0012,-0.0005]} \\ 
& {} & {0.05} & {3.03}
& {2.52} & {-0.0013} & 
{rejected} & {[-0.0012,-0.0005]} \\ 
& {} & {0.10} & {3.03}
& {2.20} & {-0.0013} & 
{rejected} & {[-0.0013,-0.0005]} \\ 
\hline
Region & {10} & {0.01} & 
{3.12} & {3.14} & {
-0.0008} & {rejected} & {
[-0.0012,-0.0008]} \\ 
& {} & {0.05} & {3.12}
& {2.62} & {-0.0009} & 
{rejected} & {[-0.0011,-0.0009]} \\ 
& {} & {0.10} & {3.12}
& {2.38} & {-0.0010} & 
{rejected} & {[-0.0011,-0.0009]} \\ 
\end{tabular}%
\caption{PCP test using the covariances for variables other than the age of
the car: neural network estimates }
\label{tab:cvalsinter:covNNother}
\end{table}

\bigskip

\begin{table}[th]
\centering%
\begin{tabular}{l|ccccccc}
 Group & {Modalities} & Test level & $k_{0}$ & $k$
& Test statistic & PCP & Confidence interval \\ 
\midrule Car group & {6} & {0.01} & 
{2.96} & {2.85} & {
-0.0070} & {rejected} & {
[-0.0069,-0.0044]} \\ 
& {} & {0.05} & {2.96}
& {2.31} & {-0.0071} & 
{rejected} & {[-0.0071,-0.0043]} \\ 
& {} & {0.10} & {2.96}
& {2.00} & {-0.0072} & 
{rejected} & {[-0.0071,-0.0042]} \\ 
\hline
Insuree age & {9} & {0.01} & 
{3.11} & {2.97} & {
-0.0067} & {rejected} & {
[-0.0067,-0.0037]} \\ 
& {} & {0.05} & {3.11}
& {2.51} & {-0.0069} & 
{rejected} & {[-0.0069,-0.0036]} \\ 
& {} & {0.10} & {3.11}
& {2.23} & {-0.0071} & 
{rejected} & {[-0.0069,-0.0035]} \\ 
\hline
Gender & {2} & {0.01} & 
{2.60} & {2.61} & {
-0.0054} & {rejected} & {
[-0.0062,-0.0054]} \\ 
& {} & {0.05} & {2.60}
& {2.00} & {-0.0055} & 
{rejected} & {[-0.0061,-0.0055]} \\ 
& {} & {0.10} & {2.60}
& {1.65} & {-0.0056} & 
{rejected} & {[-0.0061,-0.0055]} \\ 
\hline
Zone & {5} & {0.01} & 
{2.89} & {2.72} & {
-0.0068} & {rejected} & {
[-0.0067,-0.0040]} \\ 
& {} & {0.05} & {2.89}
& {2.16} & {-0.0070} & 
{rejected} & {[-0.0069,-0.0039]} \\ 
& {} & {0.10} & {2.89}
& {1.80} & {-0.0071} & 
{rejected} & {[-0.0070,-0.0039]} \\ 
\hline
Usage & {4} & {0.01} & 
{2.71} & {2.80} & {
-0.0048} & {rejected} & {
[-0.0057,-0.0048]} \\ 
& {} & {0.05} & {2.71}
& {2.25} & {-0.0050} & 
{rejected} & {[-0.0056,-0.0049]} \\ 
& {} & {0.10} & {2.71}
& {1.97} & {-0.0051} & 
{rejected} & {[-0.0056,-0.0050]} \\ 
\hline
Profession & {8} & {0.01} & 
{3.03} & {2.99} & {
-0.0062} & {rejected} & {
[-0.0060,-0.0044]} \\ 
& {} & {0.05} & {3.03}
& {2.54} & {-0.0067} & 
{rejected} & {[-0.0065,-0.0044]} \\ 
& {} & {0.10} & {3.03}
& {2.24} & {-0.0070} & 
{rejected} & {[-0.0067,-0.0043]} \\ 
\hline
Region & {10} & {0.01} & 
{3.12} & {3.14} & {
-0.0065} & {rejected} & {
[-0.0064,-0.0057]} \\ 
& {} & {0.05} & {3.12}
& {2.62} & {-0.0068} & 
{rejected} & {[-0.0067,-0.0056]} \\ 
& {} & {0.10} & {3.12}
& {2.38} & {-0.0069} & 
{rejected} & {[-0.0068,-0.0055]} 
\end{tabular}%
\caption{PCP test using the correlation coefficient for variables other than
the age of the car: neural network estimates }
\label{tab:cvalsinter:rhoNNother}
\end{table}

\bigskip

\begin{table}[th]
\centering%
\begin{tabular}{l|rrrrrlrr}
 Group & {Modalities} & Test level & $k_{0}$ & $k$
& Test statistic & PCP & \multicolumn{2}{r}{CI lowerCI upper} \\ 
\midrule Car group & {6} & 0.01 & 2.89 & 2.80 & -0.0035 & 
rejected & \multicolumn{2}{r}{[-0.0035, -0.0003]} \\ 
& {} & 0.05 & 2.89 & 2.13 & -0.0036 & rejected & 
\multicolumn{2}{r}{[-0.0036, -0.0002]} \\ 
& {} & 0.10 & 2.89 & 1.89 & -0.0037 & rejected & 
\multicolumn{2}{r}{[-0.0036, -0.0002]} \\ \hline
Insuree age & {9} & 0.01 & 3.09 & 2.93 & -0.0066 & 
rejected & \multicolumn{2}{r}{[-0.0065, 0.0031]} \\ 
& {} & 0.05 & 3.09 & 2.31 & -0.0067 & rejected & 
\multicolumn{2}{r}{[-0.0066, 0.0032]} \\ 
& {} & 0.10 & 3.09 & 2.03 & -0.0068 & rejected & 
\multicolumn{2}{r}{[-0.0067, 0.0032]} \\ \hline
Gender & {2} & 0.01 & 2.45 & 2.32 & -0.0041 & rejected & 
\multicolumn{2}{r}{[-0.0040, -0.0014]} \\ 
& {} & 0.05 & 2.45 & 1.64 & -0.0043 & rejected & 
\multicolumn{2}{r}{[-0.0042, -0.0013]} \\ 
& {} & 0.10 & 2.45 & 1.24 & -0.0044 & rejected & 
\multicolumn{2}{r}{[-0.0043, -0.0013]} \\ \hline
Zone & {5} & 0.01 & 2.80 & 2.80 & -0.0017 & rejected & 
\multicolumn{2}{r}{[-0.0016, 0.0031]} \\ 
& {} & 0.05 & 2.80 & 2.20 & -0.0017 & rejected & 
\multicolumn{2}{r}{[-0.0017, 0.0032]} \\ 
& {} & 0.10 & 2.80 & 1.92 & -0.0018 & rejected & 
\multicolumn{2}{r}{[-0.0017, 0.0032]} \\ \hline
Usage & {4} & 0.01 & 2.80 & 2.36 & -0.0060 & rejected & 
\multicolumn{2}{r}{[-0.0060, 0.0001]} \\ 
& {} & 0.05 & 2.80 & 1.68 & -0.0062 & rejected & 
\multicolumn{2}{r}{[-0.0061, 0.0001]} \\ 
& {} & 0.10 & 2.80 & 1.28 & -0.0063 & rejected & 
\multicolumn{2}{r}{[-0.0062, 0.0001]} \\ \hline
Profession & {8} & 0.01 & 3.03 & 2.80 & -0.0039 & rejected
& \multicolumn{2}{r}{[-0.0039, 0.0013]} \\ 
& {} & 0.05 & 3.03 & 2.12 & -0.0041 & rejected & 
\multicolumn{2}{r}{[-0.0040, 0.0014]} \\ 
& {} & 0.10 & 3.03 & 1.83 & -0.0041 & rejected & 
\multicolumn{2}{r}{[-0.0041, 0.0014]} \\ \hline
Region & {10} & 0.01 & 3.09 & 2.88 & -0.0042 & rejected & 
\multicolumn{2}{r}{[-0.0041, -0.0010]} \\ 
& {} & 0.05 & 3.09 & 2.43 & -0.0044 & rejected & 
\multicolumn{2}{r}{[-0.0043, -0.0009]} \\ 
& {} & 0.10 & 3.09 & 2.13 & -0.0046 & rejected & 
\multicolumn{2}{r}{[-0.0044, -0.0008]} \\ 
\end{tabular}%
\caption{PCP test using the correlation coefficient for other variables than
the age of the car: random forest estimation}
\label{tab:RFintersectiontestCOV}
\end{table}

\bigskip

\begin{table}[th]
\centering%
\begin{tabular}{l|rrrrrlr}
 Group & {Modalities} & Test level & $k_{0}$ & $k$
& Test statistic & PCP & Confidence interval \\ 
\midrule Car group & {6} & 0.01 & 2.89 & 2.75 & -0.0149 & 
rejected & [-0.0146, -0.0084] \\ 
& {} & 0.05 & 2.89 & 2.09 & -0.0155 & rejected & 
[-0.0153, -0.0082] \\ 
& {} & 0.10 & 2.89 & 1.82 & -0.0158 & rejected & 
[-0.0155, -0.0081] \\ \hline
Insuree age & {9} & 0.01 & 3.09 & 2.99 & -0.0183 & 
rejected & [-0.0182, -0.0030] \\ 
& {} & 0.05 & 3.09 & 2.41 & -0.0186 & rejected & 
[-0.0185, -0.0027] \\ 
& {} & 0.10 & 3.09 & 2.11 & -0.0188 & rejected & 
[-0.0186, -0.0025] \\ \hline
Gender & {2} & 0.01 & 2.45 & 2.32 & -0.0155 & rejected & 
[-0.0153, -0.0084] \\ 
& {} & 0.05 & 2.45 & 1.64 & -0.0159 & rejected & 
[-0.0157, -0.0083] \\ 
& {} & 0.10 & 2.45 & 1.24 & -0.0162 & rejected & 
[-0.0159, -0.0082] \\ \hline
Zone & {5} & 0.01 & 2.80 & 2.25 & -0.0165 & rejected & 
[-0.0163, -0.0059] \\ 
& {} & 0.05 & 2.80 & 1.63 & -0.0168 & rejected & 
[-0.0166, -0.0054] \\ 
& {} & 0.10 & 2.80 & 1.27 & -0.0169 & rejected & 
[-0.0168, -0.0052] \\ \hline
Usage & {4} & 0.01 & 2.80 & 2.36 & -0.0195 & rejected & 
[-0.0193, 0.0006] \\ 
& {} & 0.05 & 2.80 & 1.68 & -0.0201 & rejected & 
[-0.0198, 0.0009] \\ 
& {} & 0.10 & 2.80 & 1.28 & -0.0206 & rejected & 
[-0.0201, 0.0011] \\ \hline
Profession & {8} & 0.01 & 3.03 & 2.86 & -0.0137 & rejected
& [-0.0136, -0.0059] \\ 
& {} & 0.05 & 3.03 & 2.36 & -0.0139 & rejected & 
[-0.0138, -0.0056] \\ 
& {} & 0.10 & 3.03 & 2.11 & -0.0140 & rejected & 
[-0.0139, -0.0055] \\ \hline
Region & {10} & 0.01 & 3.09 & 2.99 & -0.0166 & rejected & 
[-0.0162, -0.0033] \\ 
& {} & 0.05 & 3.09 & 2.52 & -0.0172 & rejected & 
[-0.0169, -0.0031] \\ 
& {} & 0.10 & 3.09 & 2.23 & -0.0176 & rejected & 
[-0.0172, -0.0030] 
\end{tabular}%
\caption{PCP test using the correlation coefficient for for other variables
than the age of the car: gradient-boosted tree estimation}
\label{tab:GBTintersectiontestCOV}
\end{table}

\bigskip

\begin{table}[th]
\centering%
\begin{tabular}{r|cc|cc|cc}
Modality & \multicolumn{2}{c}{Neural network} & 
\multicolumn{2}{c}{Random forest} & \multicolumn{2}{c}{Gradient-boosted
tree} \\ 
\toprule & Raw & DD & {Raw} & DD & {Raw
} & DD \\ 
\midrule1 & {-0.0223} & {$\underset{%
(0.0004)}{-0.0053}$} & -0.0183 & $\underset{(0.0003)}{-0.0010}$ & -0.0432 & $%
\underset{(0.0008)}{-0.0139}$ \\ 
2 & {-0.0123} & {$\underset{(0.0003)}{%
-0.0058}$} & -0.0151 & $\underset{(0.0002)}{0.0002}$ & -0.0317 & $\underset{%
(0.0005)}{-0.0070}$ \\ 
3 & {-0.0112} & {$\underset{(0.0003)}{%
-0.0048}$} & -0.0209 & $\underset{(0.0003)}{-0.0033}$ & -0.0371 & $\underset{%
(0.0005)}{-0.0102}$ \\ 
4 & {-0.0024} & {$\underset{(0.0003)}{%
-0.0036}$} & -0.0227 & $\underset{(0.0003)}{-0.0033}$ & -0.0613 & $\underset{%
(0.0009)}{-0.0174}$ \\ 
5 & {-0.0143} & {$\underset{(0.0003)}{%
-0.0075}$} & -0.0142 & $\underset{(0.0002)}{-0.0036}$ & -0.0330 & $\underset{%
(0.0006)}{-0.0132}$ \\ 
6 & {-0.0155} & {$\underset{(0.0003)}{%
-0.0079}$} & -0.0158 & $\underset{(0.0002)}{-0.0040}$ & 0.0064 & $\underset{%
(0.0011)}{-0.0062}$ \\ 
\end{tabular}%
\caption{Group-averaged correlations: car group}
\label{tab:CarGroup}
\end{table}

\bigskip

\begin{table}[th]
\centering%
\begin{tabular}{r|cc|cc|cc}
\toprule Modality & \multicolumn{2}{c}{Neural network} & 
\multicolumn{2}{c}{Random forest} & \multicolumn{2}{c}{Gradient-boosted
tree} \\ 
\toprule & {Raw} & DD & {Raw} & DD & 
{Raw} & DD \\ 
\midrule0 & {-0.0079} & {$\underset{%
(0.0002)}{-0.0058}$} & -0.0271 & $\underset{(0.0003)}{-0.0048}$ & -0.0580 & $%
\underset{(0.0006)}{-0.0169}$ \\ 
1 & {-0.0152} & {$\underset{(0.0002)}{%
-0.0058}$} & -0.0130 & $\underset{(0.0000)}{-0.0012}$ & -0.0208 & $\underset{%
(0.0002)}{-0.0079}$ \\ 
\end{tabular}%
\caption{Group-averaged correlations: gender}
\label{tab:Gender}
\end{table}

\bigskip

\begin{table}[th]
\centering%
    \begin{tabular}{r|cc|cc|cc}
        \toprule Modality & \multicolumn{2}{c}{Neural network} & 
        \multicolumn{2}{c}{Random forest} & \multicolumn{2}{c}{Gradient-boosted
        tree} \\ 
\toprule & {Raw} & DD & {Raw} & DD & 
{Raw} & DD \\ 
\midrule0 & {-0.0138} & {$\underset{%
(0.0003)}{-0.0068}$} & {-0.0237} & {$%
\underset{(0.0003)}{-0.0056}$} & -0.0306 & $\underset{(0.0005)}{-0.0125}$ \\ 
1 & {-0.0150} & {$\underset{(0.0003)}{%
-0.0075}$} & {-0.0238} & {$\underset{%
(0.0003)}{-0.0073}$} & -0.0481 & $\underset{(0.0006)}{-0.0201}$ \\ 
2 & {-0.0181} & {$\underset{(0.0004)}{%
-0.0080}$} & {-0.0273} & {$\underset{%
(0.0005)}{-0.0049}$} & -0.0504 & $\underset{(0.0008)}{-0.0175}$ \\ 
3 & {-0.0176} & {$\underset{(0.0006)}{%
-0.0061}$} & {-0.0284} & {$\underset{%
(0.0004)}{-0.0037}$} & -0.0414 & $\underset{(0.0010)}{-0.0128}$ \\ 
4 & {-0.0191} & {$\underset{(0.0007)}{%
-0.0073}$} & {-0.0259} & {$\underset{%
(0.0006)}{-0.0033}$} & -0.0511 & $\underset{(0.0016)}{-0.0135}$ \\ 
5 & {-0.0172} & {$\underset{(0.0007)}{%
-0.0076}$} & {-0.0284} & {$\underset{%
(0.0006)}{-0.0029}$} & -0.0420 & $\underset{(0.0012)}{-0.0138}$ \\ 
6 & {-0.0130} & {$\underset{(0.0005)}{%
-0.0069}$} & {-0.0129} & {$\underset{%
(0.0004)}{-0.0002}$} & -0.0336 & $\underset{(0.0011)}{-0.0138}$ \\ 
7 & {-0.0076} & {$\underset{(0.0003)}{%
-0.0045}$} & {-0.0048} & {$\underset{%
(0.0001)}{0.0011}$} & -0.0049 & $\underset{(0.0007)}{-0.0014}$ \\ 
8 & {-0.0016} & {$\underset{(0.0004)}{%
-0.0026}$} & {-0.0012} & {$\underset{%
(0.0002)}{0.0034}$} & -0.0289 & $\underset{(0.0007)}{-0.0097}$ \\
\end{tabular}%
\caption{Group-averaged correlations: insuree age}
\label{tab:insuree age}
\end{table}

\bigskip

\begin{table}[th]
\centering%
\begin{tabular}{r|cc|cc|cc}
    \toprule Modality & \multicolumn{2}{c}{Neural network} & 
    \multicolumn{2}{c}{Random forest} & \multicolumn{2}{c}{Gradient-boosted
    tree} \\ 
\toprule & Raw & DD & Raw & DD & Raw & DD \\ 
\midrule1 & -0.0149 & $\underset{(0.0008)}{-0.0070}$ & -0.0310 & $\underset{%
(0.0007)}{-0.0026}$ & -0.0359 & $\underset{(0.0022)}{-0.0057}$ \\ 
2 & -0.0195 & $\underset{(0.0002)}{-0.0064}$ & -0.0305 & $\underset{(0.0002)}%
{-0.0045}$ & -0.0438 & $\underset{(0.0004)}{-0.0147}$ \\ 
3 & -0.0038 & $\underset{(0.0002)}{-0.0037}$ & -0.0038 & $\underset{(0.0001)}%
{0.0013}$ & -0.0245 & $\underset{(0.0004)}{-0.0045}$ \\ 
4 & -0.0119 & $\underset{(0.0009)}{-0.0057}$ & -0.0189 & $\underset{(0.0002)}%
{-0.0024}$ & -0.0126 & $\underset{(0.0016)}{-0.0074}$ \\ 
5 & -0.0166 & $\underset{(0.0012)}{-0.0097}$ & -0.0121 & $\underset{(0.0003)}%
{-0.0013}$ & 0.0042 & $\underset{(0.0033)}{-0.0017}$ \\ 
6 & -0.0014 & $\underset{(0.0010)}{-0.0030}$ & -0.0146 & $\underset{(0.0004)}%
{0.0016}$ & -0.0438 & $\underset{(0.0024)}{-0.0041}$ \\ 
7 & -0.0172 & $\underset{(0.0004)}{-0.0040}$ & -0.0179 & $\underset{(0.0001)}%
{-0.0022}$ & -0.0026 & $\underset{(0.0009)}{-0.0046}$ \\ 
8 & -0.0035 & $\underset{(0.0003)}{-0.0055}$ & -0.0048 & $\underset{(0.0002)}%
{0.0018}$ & -0.0483 & $\underset{(0.0011)}{-0.0115}$ \\ 
\end{tabular}%
\caption{Group-averaged correlations: profession}
\label{tab:profession}
\end{table}

\bigskip

\begin{table}[th]
\centering%
\begin{tabular}{r|cc|cc|cc}
    \toprule Modality & \multicolumn{2}{c}{Neural network} & 
    \multicolumn{2}{c}{Random forest} & \multicolumn{2}{c}{Gradient-boosted
    tree} \\ 
\toprule & {Raw} & DD & {Raw} & DD & 
{Raw} & DD \\ 
\midrule1 & {-0.0089} & {$\underset{%
(0.0005)}{-0.0048}$} & -0.0194 & $\underset{(0.0004)}{-0.0023}$ & -0.0515 & $%
\underset{(0.0013)}{-0.0206}$ \\ 
2 & {-0.0134} & {$\underset{(0.0004)}{%
-0.0074}$} & -0.0176 & $\underset{(0.0002)}{-0.0028}$ & -0.0361 & $\underset{%
(0.0009)}{-0.0158}$ \\ 
3 & {-0.0135} & {$\underset{(0.0005)}{%
-0.0045}$} & -0.0154 & $\underset{(0.0003)}{-0.0007}$ & -0.0339 & $\underset{%
(0.0011)}{-0.0187}$ \\ 
4 & {-0.0102} & {$\underset{(0.0005)}{%
-0.0054}$} & -0.0233 & $\underset{(0.0005)}{-0.0057}$ & -0.0343 & $\underset{%
(0.0008)}{-0.0118}$ \\ 
5 & {-0.0122} & {$\underset{(0.0003)}{%
-0.0059}$} & -0.0190 & $\underset{(0.0003)}{-0.0037}$ & -0.0343 & $\underset{%
(0.0006)}{-0.0148}$ \\ 
6 & {-0.0159} & {$\underset{(0.0003)}{%
-0.0060}$} & -0.0158 & $\underset{(0.0003)}{-0.0020}$ & -0.0153 & $\underset{%
(0.0005)}{-0.0059}$ \\ 
7 & {-0.0130} & {$\underset{(0.0003)}{%
-0.0047}$} & -0.0124 & $\underset{(0.0002)}{-0.0004}$ & -0.0113 & $\underset{%
(0.0004)}{-0.0021}$ \\ 
8 & {-0.0123} & {$\underset{(0.0005)}{%
-0.0081}$} & -0.0232 & $\underset{(0.0005)}{-0.0033}$ & -0.0554 & $\underset{%
(0.0013)}{-0.0093}$ \\ 
9 & {-0.0119} & {$\underset{(0.0005)}{%
-0.0045}$} & -0.0158 & $\underset{(0.0003)}{-0.0009}$ & -0.0485 & $\underset{%
(0.0011)}{-0.0143}$ \\ 
10 & {-0.0108} & {$\underset{(0.0004)}{%
-0.0073}$} & -0.0236 & $\underset{(0.0004)}{-0.0041}$ & -0.0505 & $\underset{%
(0.0010)}{-0.0126}$ \\ 
\end{tabular}%
\caption{Group-averaged correlations: region}
\label{tab:region}
\end{table}

\bigskip

\begin{table}[th]
\centering%
\begin{tabular}{r|cc|cc|cc}
    \toprule Modality & \multicolumn{2}{c}{Neural network} & 
    \multicolumn{2}{c}{Random forest} & \multicolumn{2}{c}{Gradient-boosted
    tree} \\ 
\toprule & {Raw} & DD & {Raw} & DD & 
{Raw} & DD \\ 
\midrule1 & {-0.0176} & {$\underset{%
(0.0004)}{-0.0059}$} & {-0.0263} & {$%
\underset{(0.0002)}{-0.0009}$} & -0.0183 & $\underset{(0.0004)}{-0.0055}$ \\ 
2 & {-0.0165} & {$\underset{(0.0002)}{%
-0.0052}$} & {-0.0252} & {$\underset{%
(0.0001)}{-0.0001}$} & -0.0427 & $\underset{(0.0004)}{-0.0070}$ \\ 
3 & {-0.0072} & {$\underset{(0.0002)}{%
-0.0054}$} & {-0.0026} & {$\underset{%
(0.0000)}{0.0001}$} & -0.0032 & $\underset{(0.0006)}{0.0020}$ \\ 
4 & {-0.0070} & {$\underset{(0.0003)}{%
-0.0056}$} & {-0.0159} & {$\underset{%
(0.0003)}{-0.0067}$} & -0.0590 & $\underset{(0.0010)}{-0.0218}$ \\ 
\end{tabular}%
\caption{Group-averaged correlations: usage}
\label{tab:usage}
\end{table}

\bigskip

\begin{table}[th]
\centering%
\begin{tabular}{r|cc|cc|cc}
    \toprule Modality & \multicolumn{2}{c}{Neural network} & 
    \multicolumn{2}{c}{Random forest} & \multicolumn{2}{c}{Gradient-boosted
    tree} \\ 
\toprule & {Raw} & DD & {Raw} & DD & 
{Raw} & DD \\ 
\midrule2 & {-0.0164} & {$\underset{%
(0.0002)}{-0.0044}$} & -0.0245 & $\underset{(0.0001)}{-0.0020}$ & -0.0196 & $%
\underset{(0.0003)}{-0.0105}$ \\ 
3 & {-0.0213} & {$\underset{(0.0003)}{%
-0.0077}$} & -0.0291 & $\underset{(0.0002)}{-0.0021}$ & -0.0449 & $\underset{%
(0.0005)}{-0.0175}$ \\ 
4 & {-0.0231} & {$\underset{(0.0005)}{%
-0.0047}$} & -0.0346 & $\underset{(0.0003)}{-0.0018}$ & -0.0410 & $\underset{%
(0.0006)}{-0.0110}$ \\ 
5 & {-0.0206} & {$\underset{(0.0006)}{%
-0.0037}$} & -0.0212 & $\underset{(0.0003)}{-0.0006}$ & -0.0003 & $\underset{%
(0.0006)}{-0.0041}$ \\ 
6 & {-0.0025} & {$\underset{(0.0002)}{%
-0.0035}$} & -0.0054 & $\underset{(0.0001)}{0.0034}$ & -0.0415 & $\underset{%
(0.0005)}{-0.0117}$ \\ 
\end{tabular}%
\caption{Group-averaged correlations: zone}
\label{tab:zone}
\end{table}

\end{document}